\documentclass{aa}
\usepackage{graphicx}
\usepackage{times}
\begin{document}
\title{Properties of sunspots in cycle 23} 
\subtitle{I. Dependence of brightness on  sunspot size and cycle phase}
\author{S. K. Mathew\inst{1,2}
\and V. Mart\' \i nez Pillet \inst{3}
\and S. K. Solanki\inst{1}
\and N. A. Krivova\inst{1}
}
\institute{ Max-Planck-Institute for Solar System Research, 37191 Katlenburg-Lindau, Germany\\
\email{solanki@mps.mpg.de, natalie@mps.mpg.de}
\and Udaipur Solar Observatory, P. O. Box 198, Udaipur - 313004, India\\
\email{shibu@prl.res.in}
\and Instituto de Astrof\' \i sica de Canarias, La Laguna, Tenerife, Spain\\
\email{vmp@iac.es}}
\offprints{Shibu K. Mathew\\
\email{shibu@prl.res.in}}
   \date{Received 07/09/2006; accepted 02/01/2007}
    \abstract
{}
{In this paper we investigate the dependence of umbral core  brightness, as well as 
the mean umbral and penumbral brightness on the phase of the solar cycle and on the 
size of the sunspot.}
{Albregtsen \& Maltby (1978) reported an increase in umbral core brightness from the 
early to the late phase of solar cycle from the analysis of 13 sunspots which cover 
solar cycles 20 and 21. Here we revisit this topic by analysing continuum 
images of more than 160 sunspots observed by the MDI instrument on board the 
SOHO spacecraft for the period between 1998 March to 2004 March, i.e. a sizable 
part of solar cycle 23. The advantage of this data set is its homogeneity, 
with no seeing fluctuations. A careful stray light correction, which is validated 
using the Mercury transit of 7th May, 2003, is carried out before the umbral and 
penumbral  intensities are determined. The influence of the Zeeman splitting of the 
nearby Ni~{\scriptsize I} spectral line on the measured ``continuum'' 
intensity  is also taken into account.}
{We did not observe any significant variation 
in umbral core, mean umbral and mean penumbral intensities with solar cycle, 
which is in contrast to earlier findings for the umbral 
core intensity. We do find a strong and clear dependence of the umbral brightness 
on sunspot size, however. The penumbral brightness also displays a weak dependence. 
The brightness-radius relationship has numerous implications, some of which, such as 
those for the energy transport in umbrae, are pointed out.}
{}

\authorrunning{Mathew, S. K. et al.}
\titlerunning{Umbral brightness evolution over cycle 23}
\maketitle                       
\section{Introduction}
Albregtsen \& Maltby (\cite{albregtsen1}, cf. Albregtsen \& Maltby 
\cite{albregtsen2}, Albregtsen et al. \cite{albregtsen}) reported a dependence 
of umbral core brightness on the phase of the solar cycle based on 13 sunspots 
observed at Oslo Solar Observatory. The umbral core is defined as the darkest 
part of the umbra. According to their findings, sunspots present in the early 
solar cycle are the darkest, while as the cycle progresses spots have increasingly 
brighter umbrae.  Also, the authors did not find any dependence of this relation 
on the size or the type of the sunspot. Following this  discovery Maltby et al. 
(\cite{maltby}) proposed three different semi-empirical model atmospheres 
for the umbral core, corresponding to early, middle and late phases of the 
solar cycle.
\begin{figure*}
\centering
\includegraphics[width=17cm]{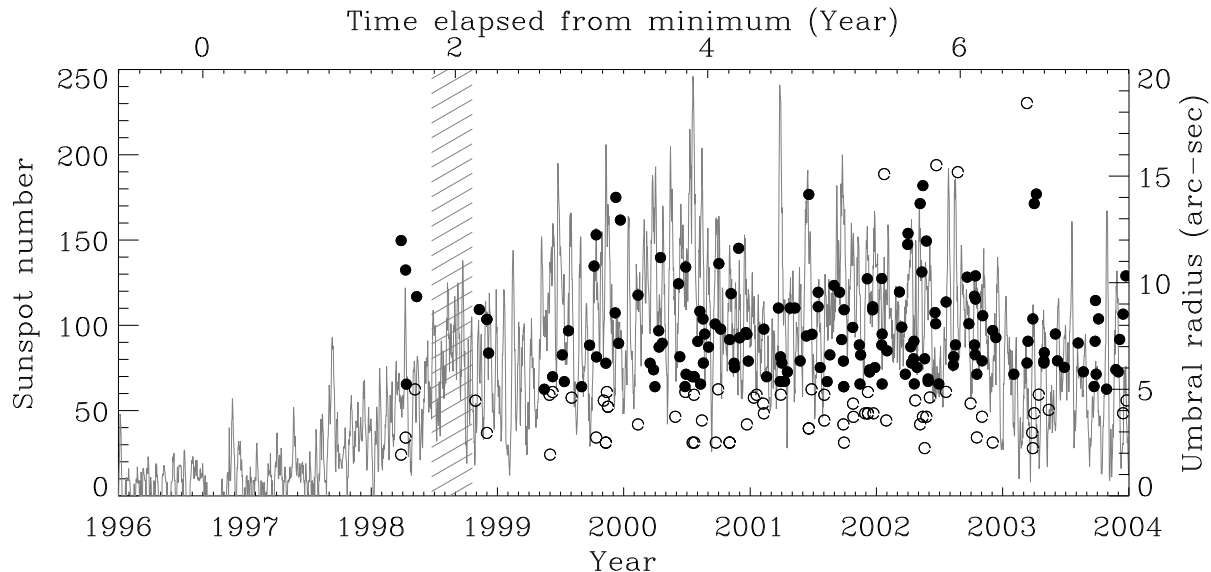} 
\caption{Distribution of analysed sunspots over the ongoing solar cycle 23. 
The grey solid line shows the International Sunspot Number (left scale). 
The solid circles show the dates of observation and umbral radii of the analysed 
sunspots with umbral radii between 5 and 15 arc-sec, and the open circles show the 
sunspots with umbral radii less than 5 arc-sec or greater than 15 arc-sec (right-hand scale). 
No sunspot in 1996 \& 1997 fulfilled the selection criteria of $\mu > 0.94$. 
The hatched area marks the period when contact with SOHO was lost.}
\end{figure*}
 
In order to explain the umbral brightness variation with solar cycle  
two hypotheses have been put forward. Sch$\mathrm{\ddot{u}}$ssler 
(\cite{schussler}) proposed that umbral brightness may be influenced 
by the age of the sub-photospheric flux tubes, whereas Yoshimura 
(\cite{yoshimura}) suggested that the brightness of the umbra 
depends on the depth in the convection zone at which the flux 
tube is formed. Confirmation of these results appears important for two reasons. 
Firstly, this is the only strong evidence for a dependence of local properties
of the magnetic features on the global cycle. E. g., the facular contrast does not 
depend on solar cycle phase (Ortiz et al. \cite{ortiz}). Secondly, such a confirmation 
appears timely in  the light of the recent paper by Norton \& Gilman 
(\cite{norton}), who reported a smooth decrease in umbral brightness from 
early to mid phase in solar cycle 23, reaching a minimum intensity around 
solar maximum, after which the umbral brightness increased again, based on the 
analysis of more than 650 sunspots observed with  the MDI instrument. 
This decrease in brightness contradicts the results of Maltby et al. (\cite {maltby}). 
Also, the data used by Norton \& Gilman (\cite{norton}) were not corrected for stray 
light and no sunspot size dependence of the brightness was discussed. 

In this paper we investigate the dependence of umbral core  brightness, 
as well as the mean umbral and penumbral brightness on the solar cycle and on the 
size of the sunspot.  In the following 
section we describe the data selection. In the third 
section we deal with the data correction for stray light and for the 
influence of Zeeman splitting of the nearby Ni~{\scriptsize I} absorption line on 
continuum measurements. In sections 4 and 5 we present our results. We discuss our results 
and compare them with earlier findings in section 6.
\section{Data selection}
Continuum full disk images recorded by the Michelson Doppler Imager (MDI; 
Scherrer et al. \cite{scherrer}) on board the SOHO spacecraft are used in 
this analysis.  The continuum images are obtained from five filtergrams 
observed around the Ni~{\scriptsize I} 6768 \AA\/ mid-photospheric 
absorption line with a spectral pass band of 94 m\AA\/ each. The filtergrams 
are summed in such a way  as to obtain the continuum intensity 
is free of Doppler cross talk at the 0.2\% level. The 
advantage of this data set is its homogeneity with no seeing fluctuations.  

We selected 234 sunspots observed between March, 1998 and March, 2004. 
The selected sunspots  were located  close to the 
disk centre, i.e. for $\mu > 0.94$, where $\mu = \cos \theta $ and $\theta$ is 
the angle between the line-of-sight and the surface normal. Also, our analysis was 
mostly restricted to regular sunspots, this excluded complex sunspots having very 
irregular shape and multiple umbrae. By looking through the daily images and selecting 
the  sunspot when it was very close to the central meridian, we make sure that a 
particular sunspot is included only once in our analysis during one solar rotation. 
Out of the selected sunspots, 164 sunspots have an umbral radius between  5\arcsec\/ 
and 15\arcsec. Even though all the 234 sunspots were used for the study of 
radius-brightness dependence, only the sunspots with umbral radius between  5\arcsec\/ 
to 15\arcsec\/ were used for the study of brightness dependence on solar cycle. 
This is done in order to facilitate a direct comparison of our results with those 
of Maltby et al. (\cite{maltby}).  The data set covers most of solar cycle 23, 
although it does miss a few sunspots in the beginning of the cycle due to our 
selection criteria and  the end of the cycle. 

Figure 1 shows the distribution of sunspots\footnote{Sunspot numbers are 
compiled by the Solar Influences Data Analysis Center 
(http://sidc.oma.be), Belgium} over cycle 23, 
used for the study of the solar cycle dependence of brightness (filled 
circles) and those used only to determine the dependence on size (open circles).  
By restricting the analysis to sunspots near the disk 
centre, we avoid the suspected effect of centre-to-limb variation 
on the umbral brightness (Albregtsen et al. \cite{albregtsen}). 
This is shown in Sect. 5. 

The determination of umbral-penumbral and penumbral-quiet 
sun boundaries was carried out using the cumulative histogram 
(Pettauer \& Brandt \cite{pettauer}) 
of the intensity of the sunspot brightness and of the immediately surrounding quiet Sun 
(whose average is set to unity). This histogram was computed for  88 
symmetric sunspots scattered across the observational period and then 
averaged. This averaged cumulative histogram is shown in Fig. 2. Note that the 
histogram is computed after stray light correction (see Sect. 3). The quiet 
Sun corresponds to the steep rise around normalised intensity unity. The rise 
at around 0.6 corresponds to the penumbra, below that is the umbra. In order to
determine the intensity threshold corresponding to the penumbra-photosphere and 
umbra-penumbra boundaries linear fits to the flattest parts of the averaged histogram 
were computed. The boundaries were chosen at the highest intensity    
at which the linear fit ceases to be a tangent to the histogram. 
The reason why the penumbra is visible as a reasonably sharp drop, while the 
umbra is not, is only partly due to the larger 
range of intensity found in the umbra. It is mainly due to the large difference in 
umbral brightness from spot to spot (see Sect. 4). From the average cumulative histogram
it was found that values of 0.655 and 0.945 in normalised 
intensity correspond to umbral-penumbral, and penumbral-quiet sun boundaries, respectively. 
These values were later used to determine the umbral and spot radius. All intensities are
normalised to quiet Sun values at roughly the same $\mu$ value as the sunspot.
\begin{figure}
\resizebox{\hsize}{!}{\includegraphics{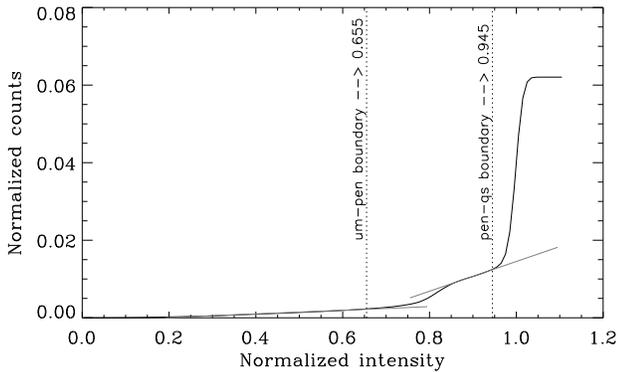}} 
\caption{Average cumulative intensity histogram used for obtaining the umbral-penumbral 
and penumbral-quiet Sun sunspot boundaries. Dotted lines are linear fits to the flattest 
parts of the histogram. Vertical dotted lines mark the values selected for umbral-penumbral 
and penumbral-quiet Sun boundaries.}
\end{figure}
\begin{figure}
\resizebox{\hsize}{!}{\includegraphics{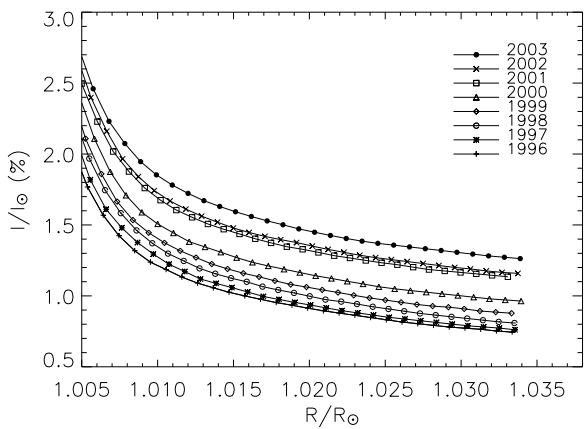}} 
\caption{Off-limb stray light profiles of SoHO/MDI continuum images for 8 different years.}
\end{figure}
\begin{figure}
\resizebox{\hsize}{!}{\includegraphics{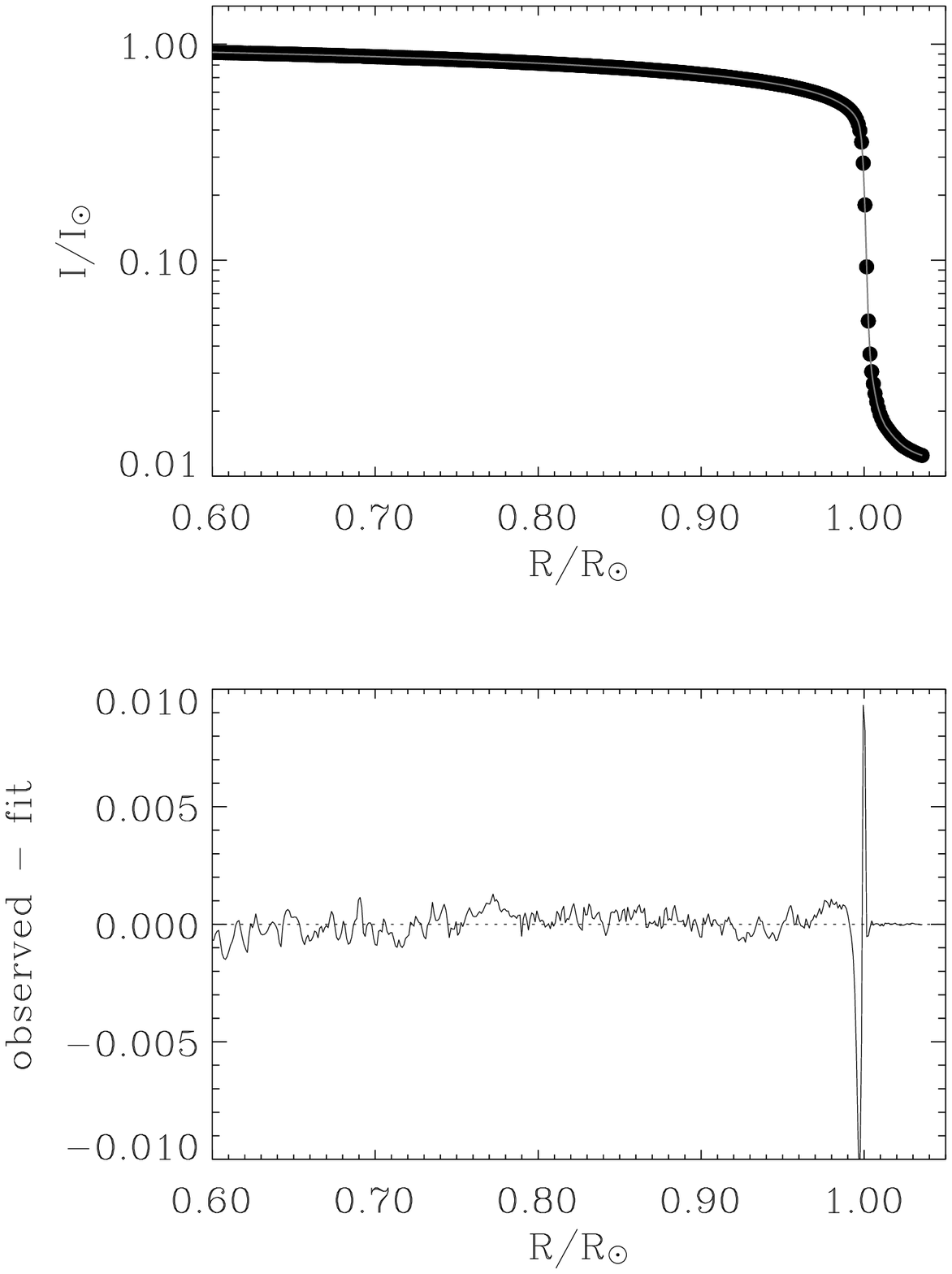}}  
\caption{A typical stray-light fit to the observed limb profile. {\bf Top:} the filled circles show the 
observed average limb profile and the solid line shows the fit to the observed profile. 
{\bf Bottom:} the residual after the fit.}
\end{figure}
\section{Data correction}
Before retrieving the brightness, we made a few  corrections to the observed data.
Even though the atmospheric seeing related blurring and distortions are absent, 
MDI continuum images are found to be contaminated by instrumental scattered light. 
By checking the falloff of  intensity just outside the solar limb, it was noticed 
that the instrumental scattered light increased with the aging of the instrument, 
which we carefully correct for.

Also, continuum measurements in MDI are not carried out  in a pure continuum spectral 
band. As described in the previous section, five filtergrams obtained around a 
Ni~{\scriptsize I} absorption line are used to compute the continuum intensities. 
We investigate the effect of this absorption line in the vicinity of a filter 
pass-band on observed continuum intensities. This is especially important in 
sunspots where the line profile changes due to the presence of a magnetic field. 
In the following subsections we elaborate on these corrections.

\subsection{Stray light correction}
In order to remove the stray light, average radial profiles were obtained 
from the observed full-disk MDI continuum images. Figure 3 shows such intensity 
profiles obtained from an MDI continuum image, averaged over the whole limb, 
each year just outside the solar disk. The gradual increase   in scattered light 
with time is clearly evident in the plot. These profiles were fitted to retrieve 
the PSF (point-spread function) of the instrument 
(Mart\' \i nez Pillet \cite{valentin}, Walton \& Preminger \cite {walton}). The 
radial profiles were generated using the spread function along with 
the centre-to-limb variation (CLV). A fifth order polynomial is used to describe
the CLV. The initial values of the CLV coefficients were taken from 
Pierce \& Slaughter (\cite {pierce}). The computed profiles were iteratively 
fitted to the observations by adjusting the coefficients of the PSF and CLV. 
A deconvolution of the observed image with the model PSF (generated from the 
fitted coefficients) is carried out to retrieve the original intensity. 

Figure 4 shows a typical fit to the observed radial profile  and the difference 
between observed and fitted profiles. The spikes in the residual are due to the 
sharp change in intensity at the solar limb and are restricted to the points just 
outside and inside the Sun. Excluding these points, the residuals always lie 
between $\pm 0.002$. 

We tested our fitting procedure for the stray light correction 
using MDI continuum images of the Mercury transit on 7th May, 2003. 
Figure 5 shows the observed and restored intensity for a 
cut across the solar disk through the Mercury image (at $\mu=0.65$) whose expected full width 
at half maximum is 12\arcsec, i.e. typical of the diameter of a sunspot umbra. 
It is evident from the figure that while the intensity in the original cut 
through the Mercury image never drops below 16\%, after the stray light removal 
the intensity drops to very close to zero. More details of the stray light  
correction are given in Appendix A. \\
\begin{figure}
\resizebox{\hsize}{!}{\includegraphics{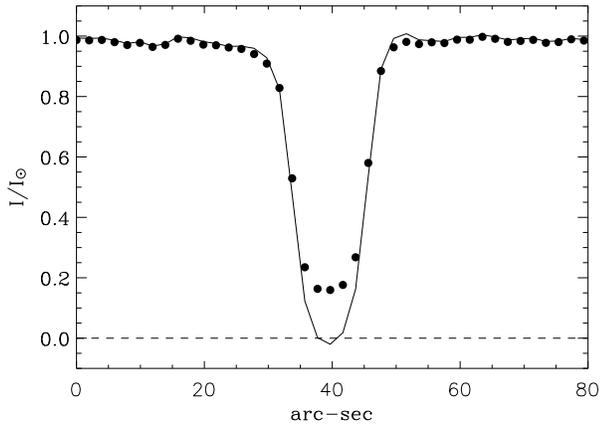}} 
\caption{Observed and restored intensity for a 
cut across the solar disk through the Mercury image.
Filled circles show the observed intensity profile through the centre of the mercury image. 
The solid line shows the profile after the stray light removal.}
\end{figure}

\subsection{Correction for the Zeeman splitting of the Ni~{\scriptsize {\rm \it I}} line} 
In MDI, the continuum is computed by measuring  intensities at five filter 
positions (designated as F$_{0}$ through F$_{4}$, cf. Scherrer et al. \cite{scherrer})  
on a spectral band which includes the Ni~{\scriptsize I} absorption line. 
The filter F$_{0}$ (whose profile is shown in Fig. 6) gives the main contribution to the
measured continuum, while the intensities recorded through the other filters are used to
correct for intrusions of the Ni~{\scriptsize I} line into the continuum filter, 
mainly introduced by Doppler shifts. The claimed accuracy for such corrections is 0.2\% 
of the continuum intensity (Scherrer et al. \cite{scherrer}), which is perfectly 
adequate for our analysis. 
\begin{figure}
\resizebox{\hsize}{!}{\includegraphics{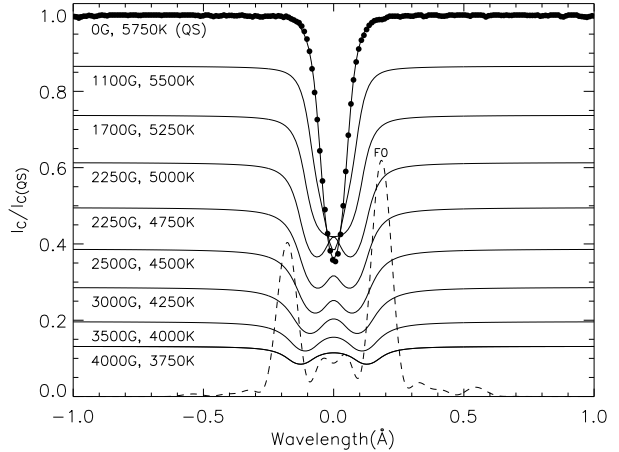}} 
\caption{Synthesised Ni~{\scriptsize I} line profiles (solid lines). 
Filled circles represent the FTS quiet Sun (QS) spectrum for the same line. The dashed curve 
shows the position of the MDI-F$_{0}$ filter transmission profile.} 
\end{figure}
\begin{figure}
\resizebox{\hsize}{!}{\includegraphics{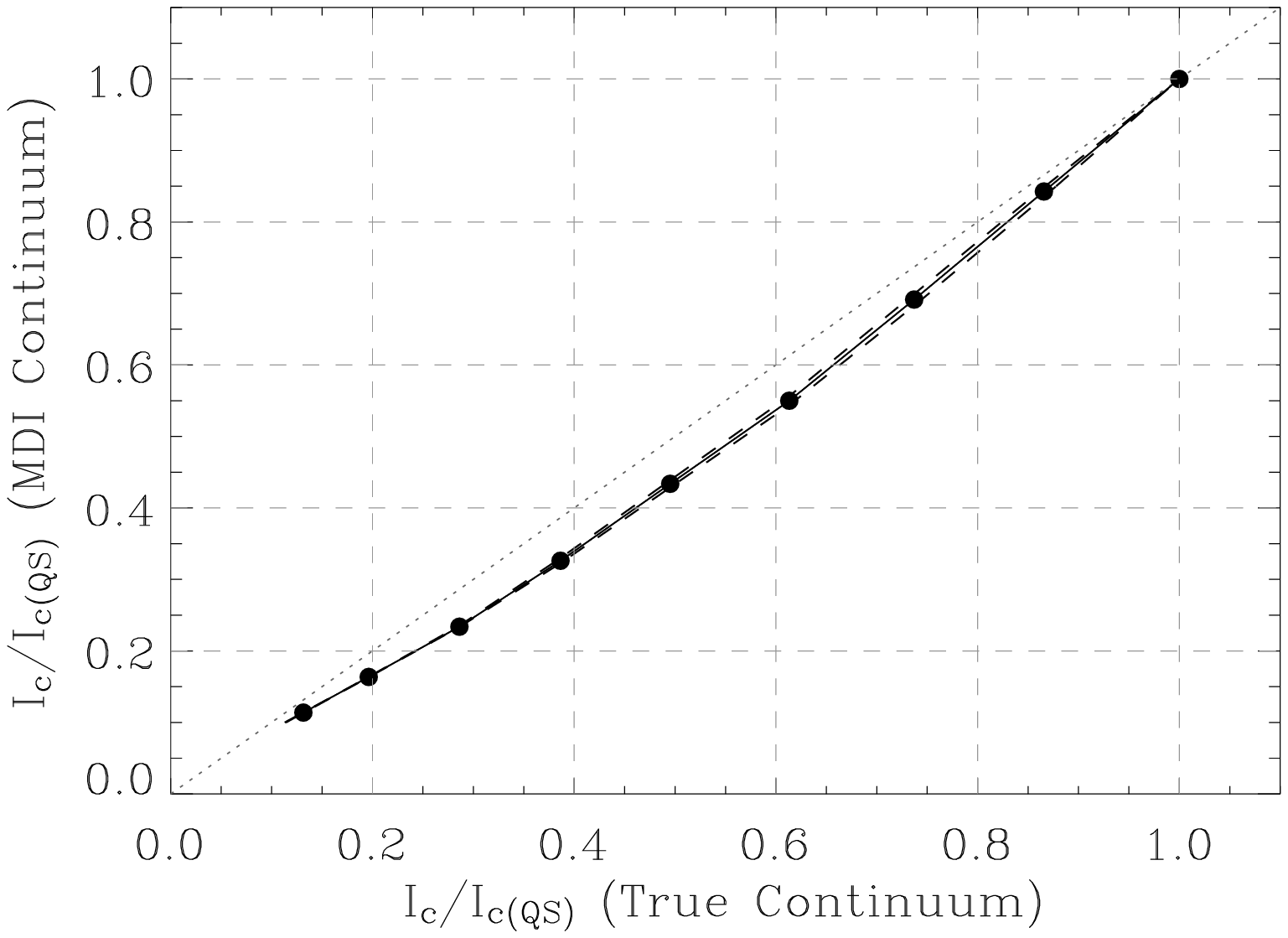}} 
\caption{Brightness correction for the contribution  of the Ni~{\scriptsize I} 
line to the MDI-continuum measurements. The solid line shows the normalised continuum 
intensity computed including the influence of the spectral line plotted versus the 
computed true continuum for the same set of model atmospheres. Dashed lines are 
for  magnetic field strengths around 10\% below or above the values listed in Table 1, 
for a given value of temperature.} 
\end{figure}

It is, however, not clear to what extent the Zeeman splitting of the line, as known to be
present in sunspots, affects the continuum intensity measurement. In order to quantify 
this effect in sunspots, we carried out a series of calculations taking various values 
for magnetic field strength and temperature, which approximately simulate the 
relationship between the magnetic field and corresponding temperatures found in  
sunspots following Kopp \& Rabin (\cite {kopp}), Solanki et al. (\cite{sol1993}), 
and Mathew et al. (\cite{mathew}). The exact choice of the field 
strength-temperature relation is not very critical, as we have found by considering 
also other combinations (e.g. with lower or with higher field strength for a 
given temperature). The magnetic field strength and the corresponding temperature 
along with other  parameters used in the synthesis of Ni~{\scriptsize I} line profiles 
are given in Table 1. The abundance is given on a logarithmic scale on which the 
hydrogen abundance is 12 and the oscillator strength implies the $\log (gf)$ value.  
\begin{table}\caption{Parameters used for producing Ni~{\scriptsize I} line profiles}
\begin{center}
\begin{tabular}{ll}\hline \hline
Centre wavelength 			& 	6767.768 \AA \\
Abundance (Ni)				&	6.25\\
Oscillator strength 			&	$-$1.84\\
Macro turbulence 			&	1.04 km s$^{-1}$\\
Micro turbulence			&       0.13 km s$^{-1}$\\
\hline
Temperature (K) & Field strength (G) \\
\hline
5750 	 	&	0    \\
5500  		&	1100 \\
5250 		&	1700 \\
5000 		&	2250 \\
4750 		&	2250 \\
4500 		&	2500 \\
4250 		&	3000 \\
4000		&	3500 \\
3750 		&	4000 \\
\hline \hline
\end{tabular}
\end{center}
\end{table}
\begin{figure}
\resizebox{\hsize}{!}{\includegraphics{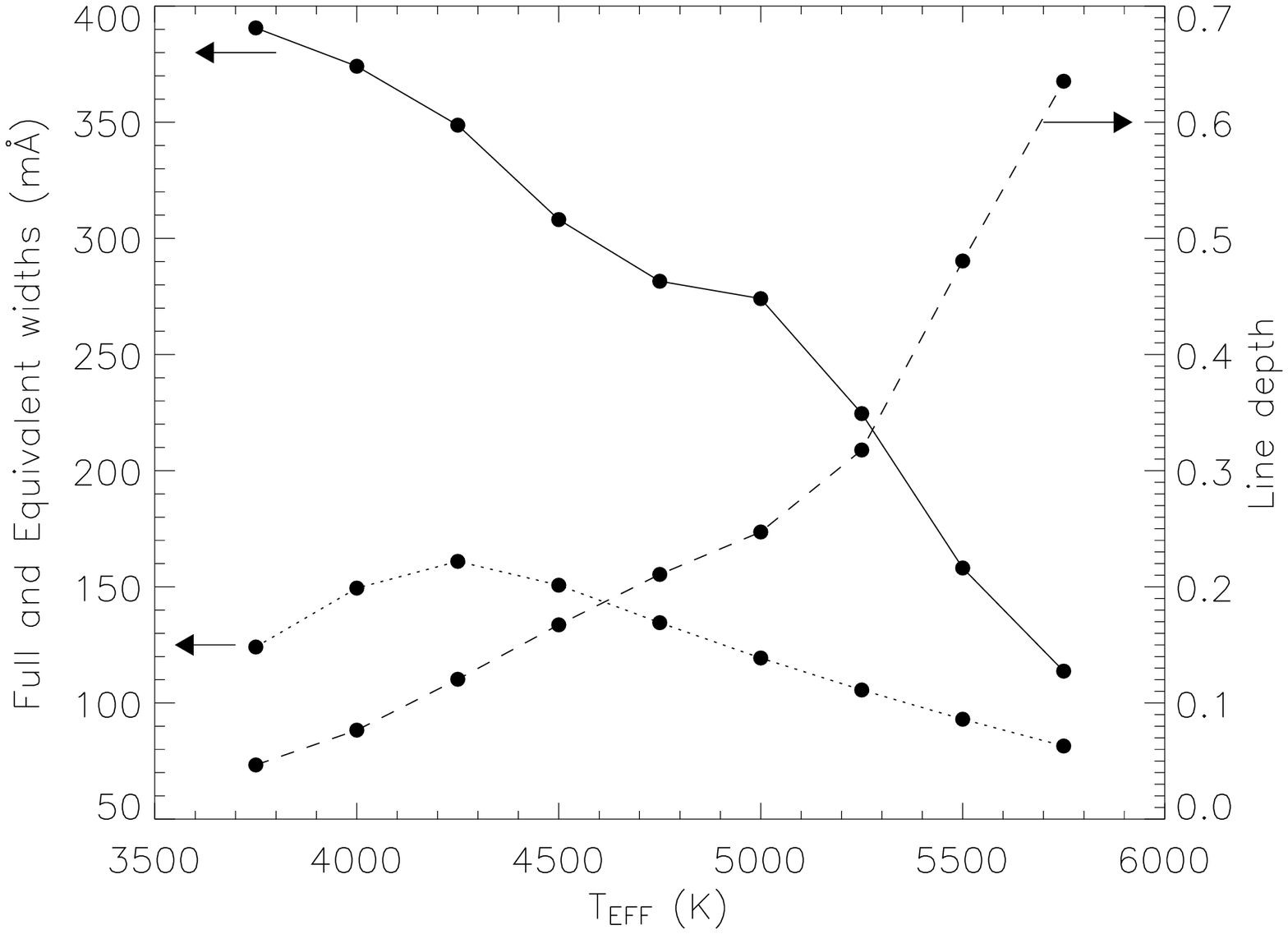}} 
\caption{The change in full width (solid line; left scale), equivalent width (dotted line; left scale) and 
line depth (dashed line; right scale) 
with effective temperature.} 
\end{figure}

Figure 6 shows the computed Ni~{\scriptsize I} line profiles along with the position 
of the MDI-F$_{0}$ filter transmission profile. The solid circles are the quiet Sun 
FTS (Fourier Transform Spectrometer) spectrum for the same line 
(Kurucz et al. \cite{kurucz_at}). Each plotted line profile was computed using a 
model atmosphere from Kurucz (\cite{kurucz}) with effective temperature and 
height independent vertical magnetic field of the strength  listed in Table 1. 
The Ni abundance is taken from Grevesse \& Sauval (\cite {grevesse}) and the atomic 
parameters are obtained from the Kurucz/NIST/VALD \footnote {Kurucz data base - http://www.pmp.uni-hannover.de \\
NIST - http://physics.nist.gov/PhysRefData\\
VALD - http://www.atro.uu.se/ $\tilde{}$~vald }  atomic data bases. 
As a first step, the Ni~{\scriptsize I} line profile taken from the quiet Sun FTS spectrum 
is fitted using the Kurucz quiet Sun model atmosphere (T$_{\rm eff} = $ 5750 K) keeping the 
micro- and macro-turbulence and oscillator strength as free parameters. The values 
obtained for the oscillator strength and micro- and macro-turbulence from the fit 
are maintained when computing all the remaining line profiles with various magnetic 
field strengths.  MDI theoretical filter transmission profiles are created for the 
five filter positions across the spectral line. The transmitted intensity through 
each filter is computed  and combined following Scherrer et al. (\cite{scherrer}) to 
derive the continuum intensities. In Fig. 7 we plot the `true continuum' intensities 
(i.e., the intensity which would have been measured through the filter if the line 
were not present in the vicinity of the MDI filter) versus the intensities resulting 
from the MDI continuum measurements for different effective temperatures. 
Clearly, the MDI continuum measurements in the presence of the Ni~{\scriptsize I} absorption  
line provide a lower intensity than the real continuum in the sunspots, and the difference 
varies with the changing field strength and temperature. Naively one would expect this
difference to increase for increasing magnetic field strengths, whereas the 
actually found behaviour is more complex. An explanation for this is given 
in Fig. 8. The decreasing temperature reduces  the line depth (which is given in units 
relative to the continuum intensity for the relevant line profile), while the 
equivalent width initially increases before decreasing again with decreasing temperature 
if the field is left unchanged.  Increasing field strength leads to enhanced line 
broadening and a slight increase in equivalent width. The combined influence of both 
effects is plotted in Fig. 8. Note that the width is measured as the wavelength 
difference between the two outer parts of the line profile at which it drops to $1-d/2$, 
where   $d$ is the line depth in units of the continuum intensity. The behaviour seen in Fig. 7 therefore 
partly reflects the dependence of the equivalent width on temperature, but quite significantly also the 
fact that the total intensity absorbed by the line in terms of the {\it continuum intensity of the quiet Sun}
decreases rapidly with decreasing temperature. This is the quality more relevant for our needs, rather than 
total absorbed intensity relative to sunspot continuum intensity, which corresponds to the equivalent width.
\begin{figure*}
\centering
\includegraphics[width=8.5cm]{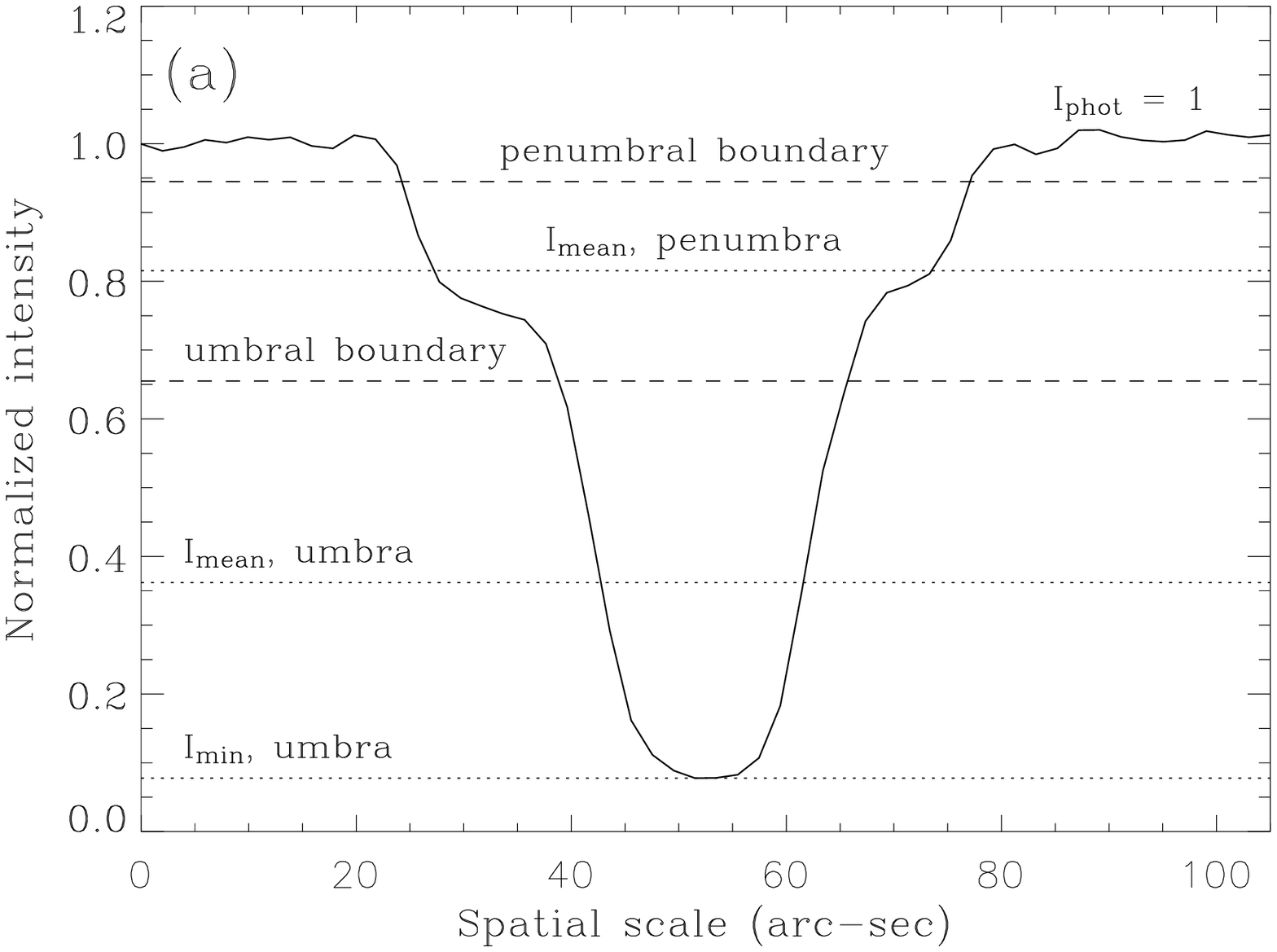} \includegraphics[width=8.5cm]{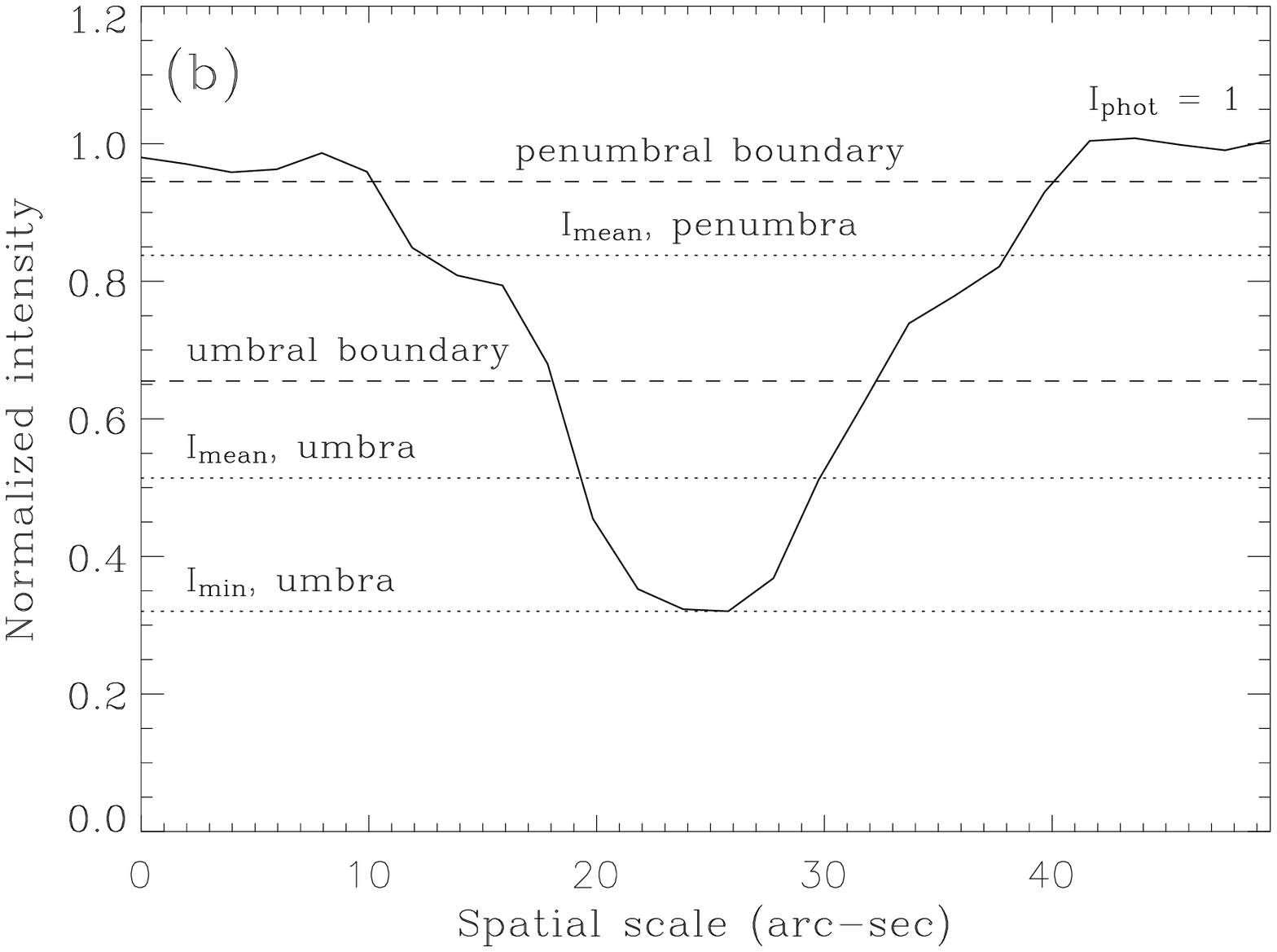}
\caption{Cut through two simple sunspots with different effective umbral radii of around 
{\bf (a)} 15 arc-sec and 
{\bf (b)} 6 arc-sec.}
\end{figure*}
\begin{figure*}
\centering
\includegraphics[width=5.6666cm]{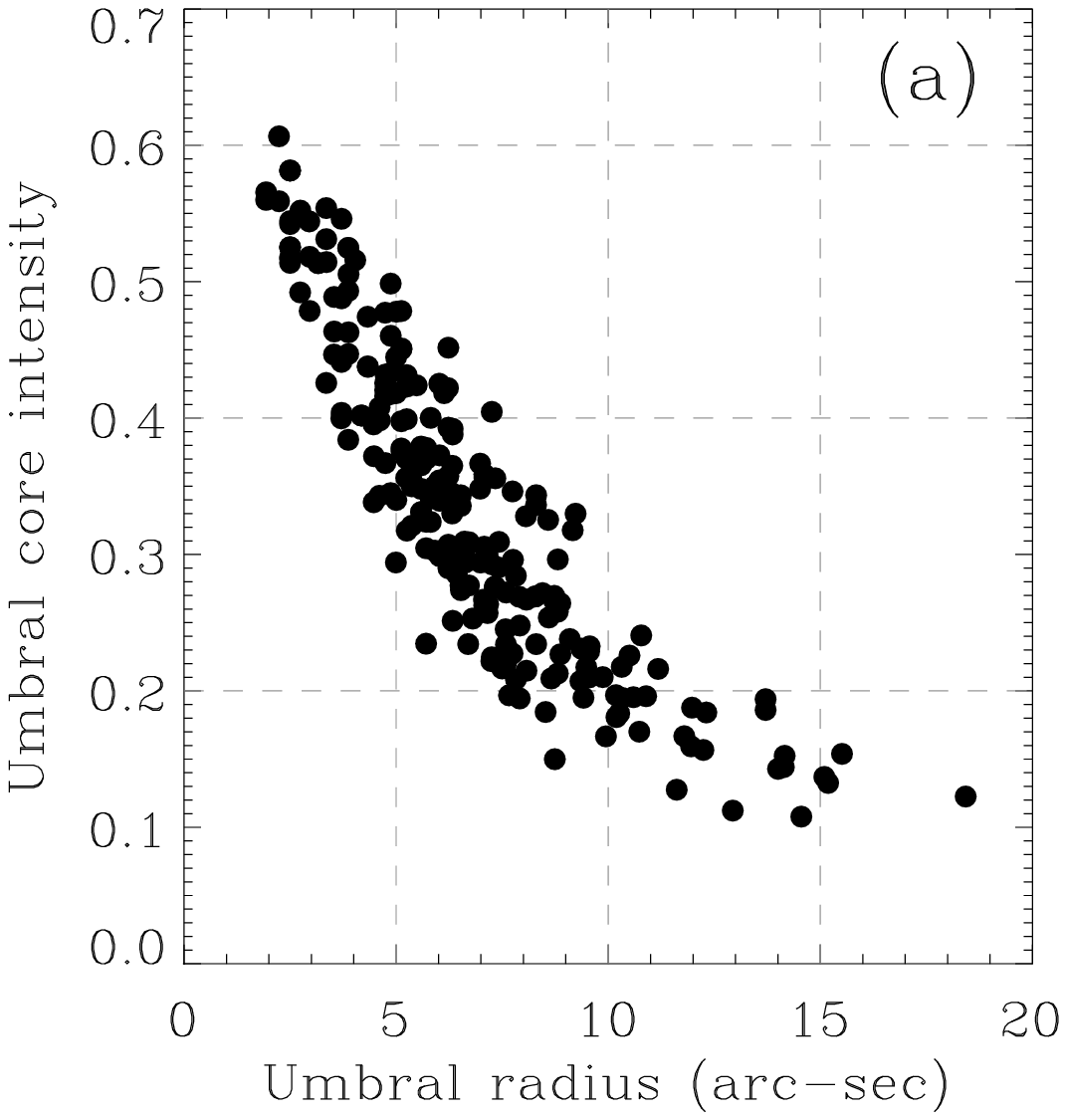} 
\includegraphics[width=5.6666cm]{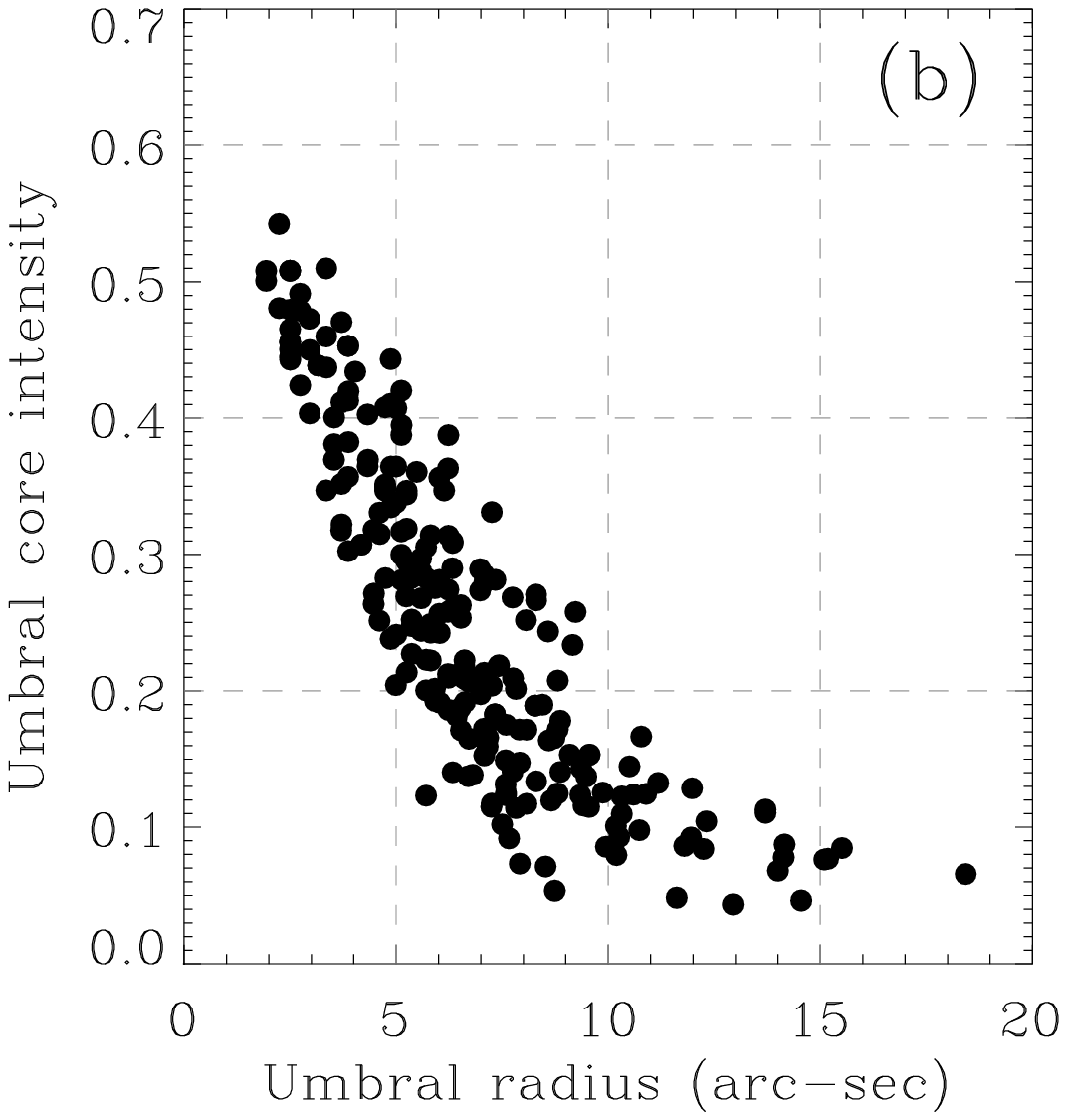}
\includegraphics[width=5.6666cm]{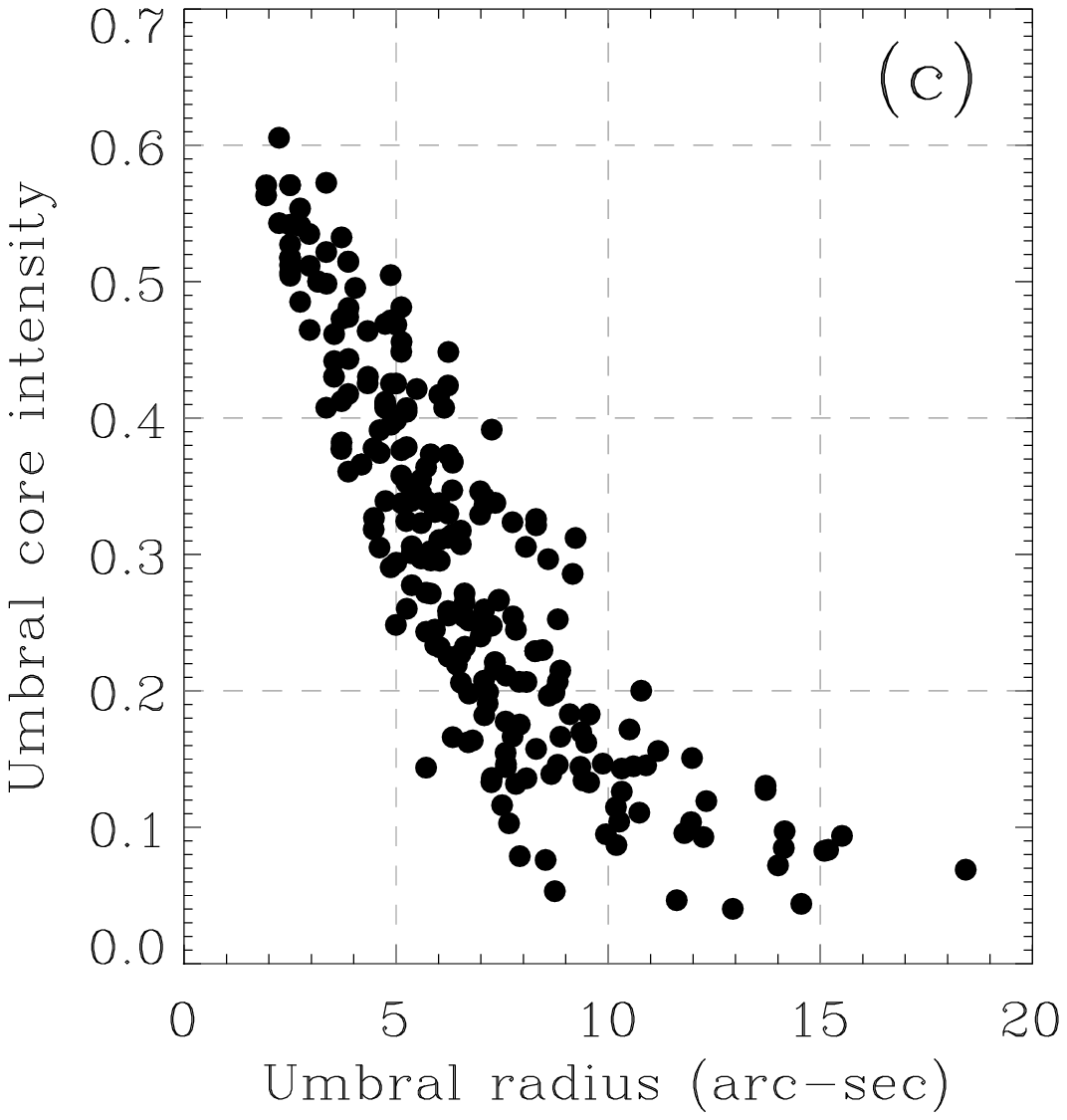}
\caption{Umbral core intensity versus umbral radius, {\bf (a)} observed, {\bf (b)} 
corrected for stray light, and {\bf (c)} corrected for stray light and the influence of 
the Ni~{\scriptsize I} \/line. Here sunspots with umbral radius less 
than 5 arc-sec and greater than 15 arc-sec are also included.}
\end{figure*}

The two dashed lines in Fig. 7, which are  hardly distinguishable from the solid line correspond to using 
different field strength-temperature relations. The upper one is found if we decrease the field strength by around 
10\% (i.e. the correction is minutely smaller), the lower one for around 10\% higher field. We use the solid line in 
Fig. 7 to  correct the observed continuum intensities in sunspots. During the data reduction process 
all the resulting intensities after the stray light removal are replaced by reading 
out the corresponding value from the computed true continuum.    

\section{Brightness-radius relationships} 
Before discussing the brightness-radius relationship,in Fig. 9 we show cuts through two different sunspots. 
Those allow us to point out various intensity values used in our study. The big sunspot (Fig. 9(a)) has an 
effective umbral  radius of around 15\arcsec\/. The horizontal dashed lines indicate the umbral and penumbral 
boundaries, whereas the dotted lines represent mean and minimum umbral intensities as well as mean penumbral 
intensity in this particular sunspot. Similarly, Fig. 9(b) shows a cut through a small sunspot (umbral 
radius $\approx 6$\arcsec). 

\subsection{Umbral core and mean intensity versus umbral radius}
Figure 10 shows the relation between umbral core intensity and umbral 
radius. The umbral radius is computed as the radius of a circle with the 
same area as the (irregularly shaped) umbra under study. 
The umbral core intensity is the lowest intensity value found in the particular 
umbra (see Fig. 9). 

All intensities are normalised to the average local quiet Sun intensity. 
Figure 10(a) shows this relation for the observed intensity and (b) for the stray light 
corrected intensities. The influence of the Ni~{\scriptsize I} 
line on the continuum measurement is still present in this panel. Figure 10(c) is corrected for 
both the stray light and the effect of the Ni~{\scriptsize I} line 
on the continuum measurement.  In all 
the figures the trend remains the same.  It is clear from the figure 
that the core intensity decreases very strongly with increasing umbral radius. 
\begin{figure}
\resizebox{\hsize}{!}{\includegraphics{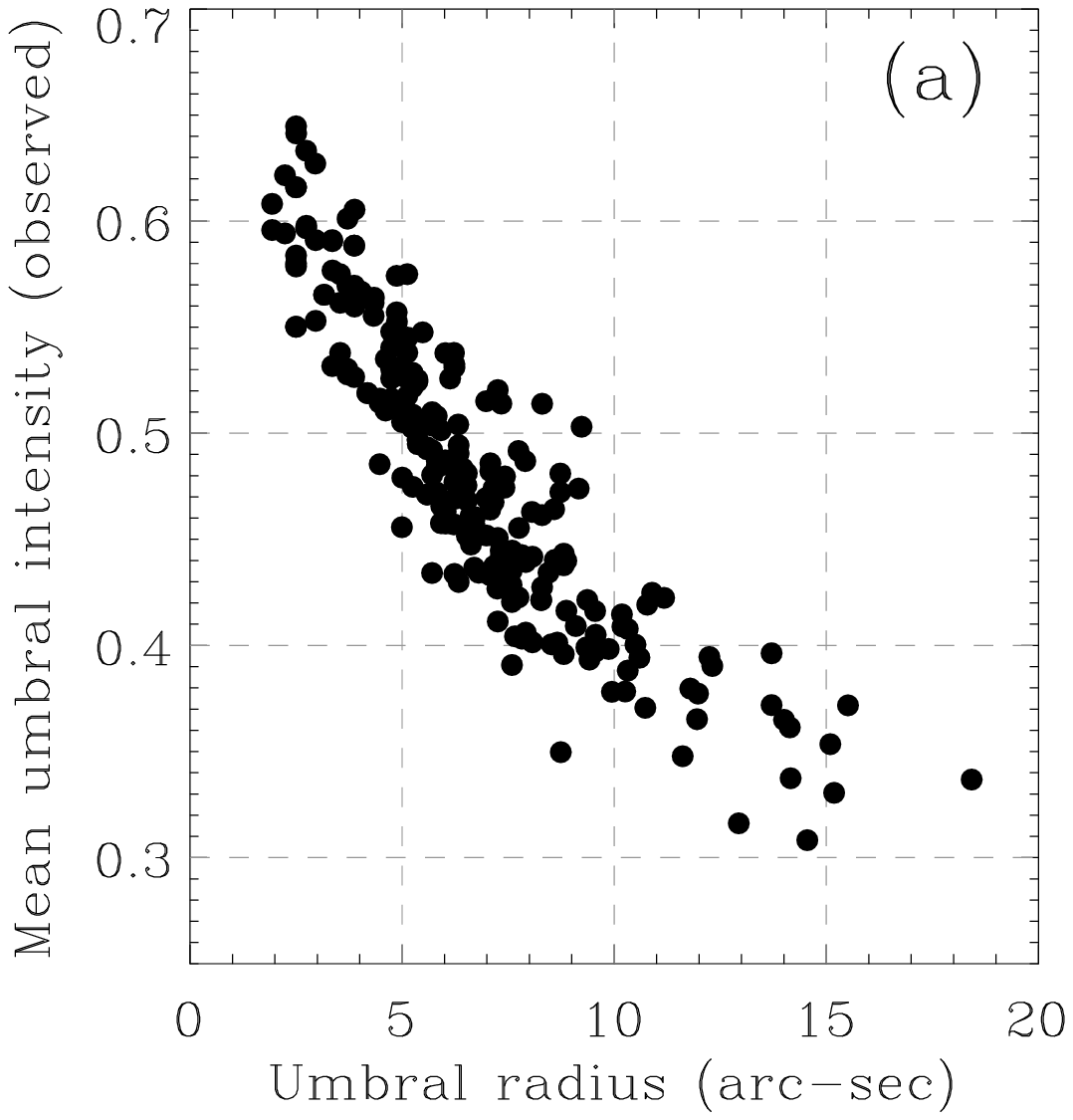}\includegraphics{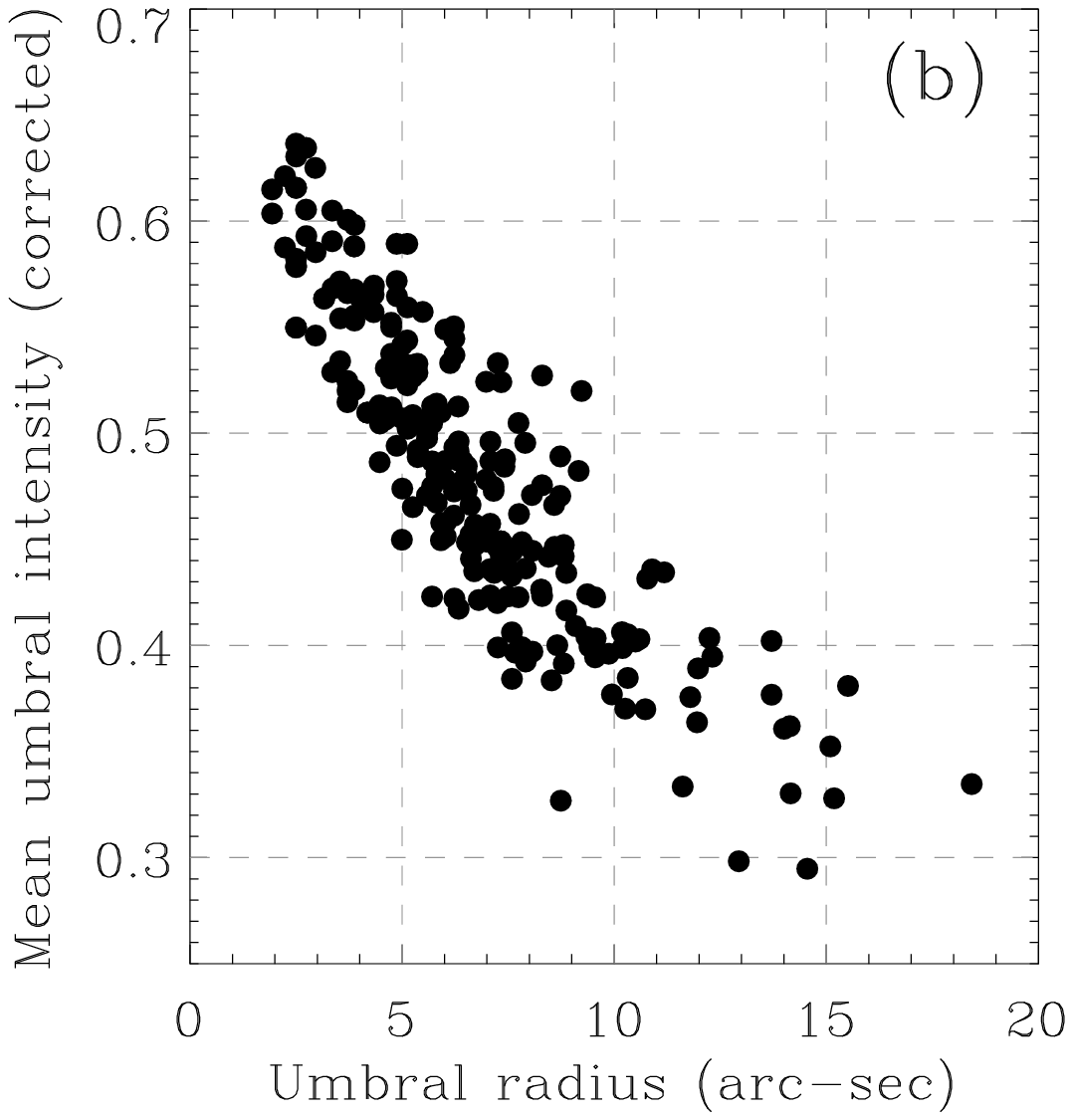}}
\caption{Mean umbral intensity versus umbral radius, {\bf (a)} observed  and, {\bf (b)} 
corrected for stray light and the influence of the Ni~{\scriptsize I} \/line. 
Here sunspots with umbral radius less than 5 arc-sec  and greater than 15 arc-sec are 
also included.}
\end{figure}
\begin{figure}
\resizebox{\hsize}{!}{\includegraphics{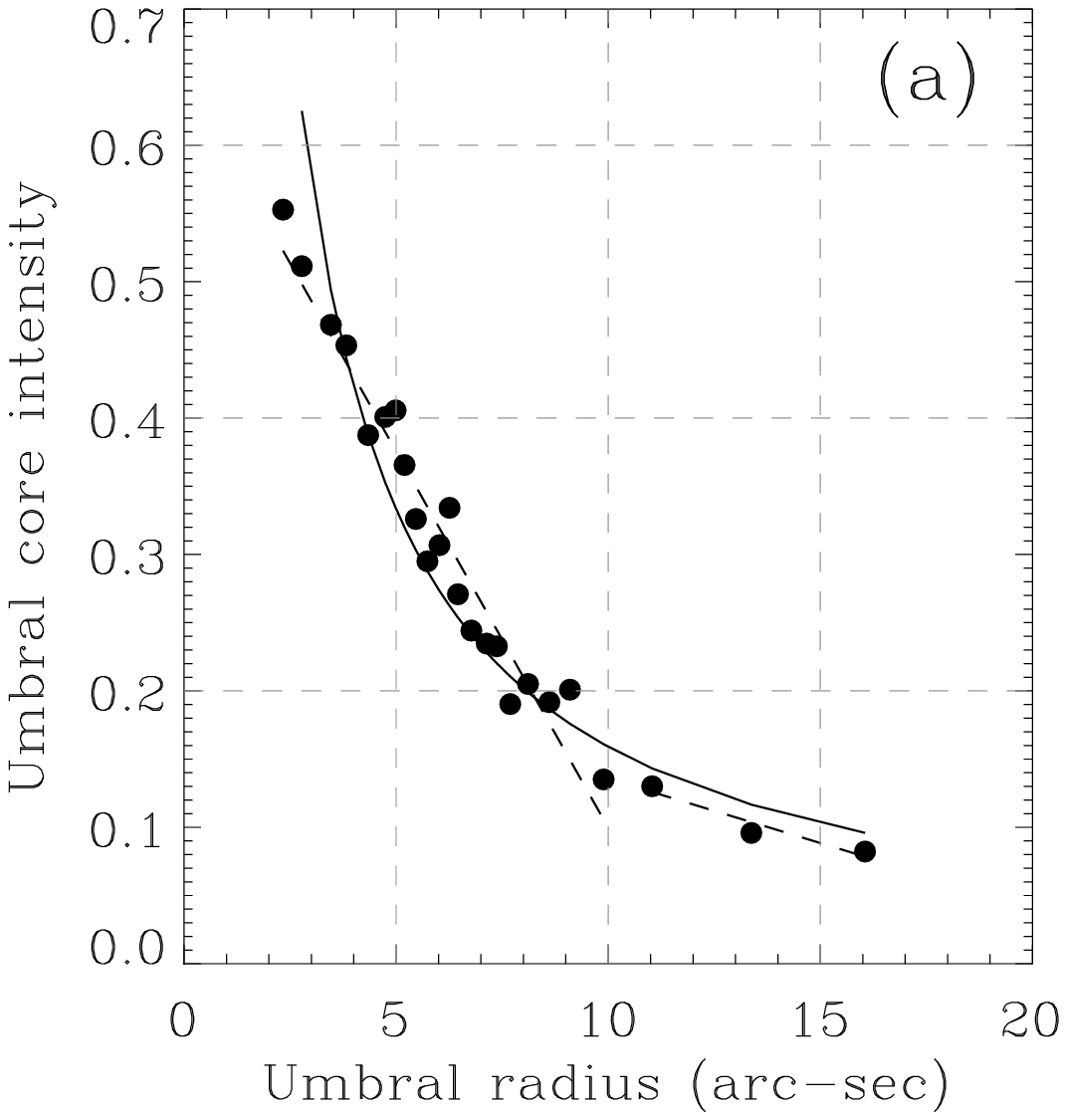}\includegraphics{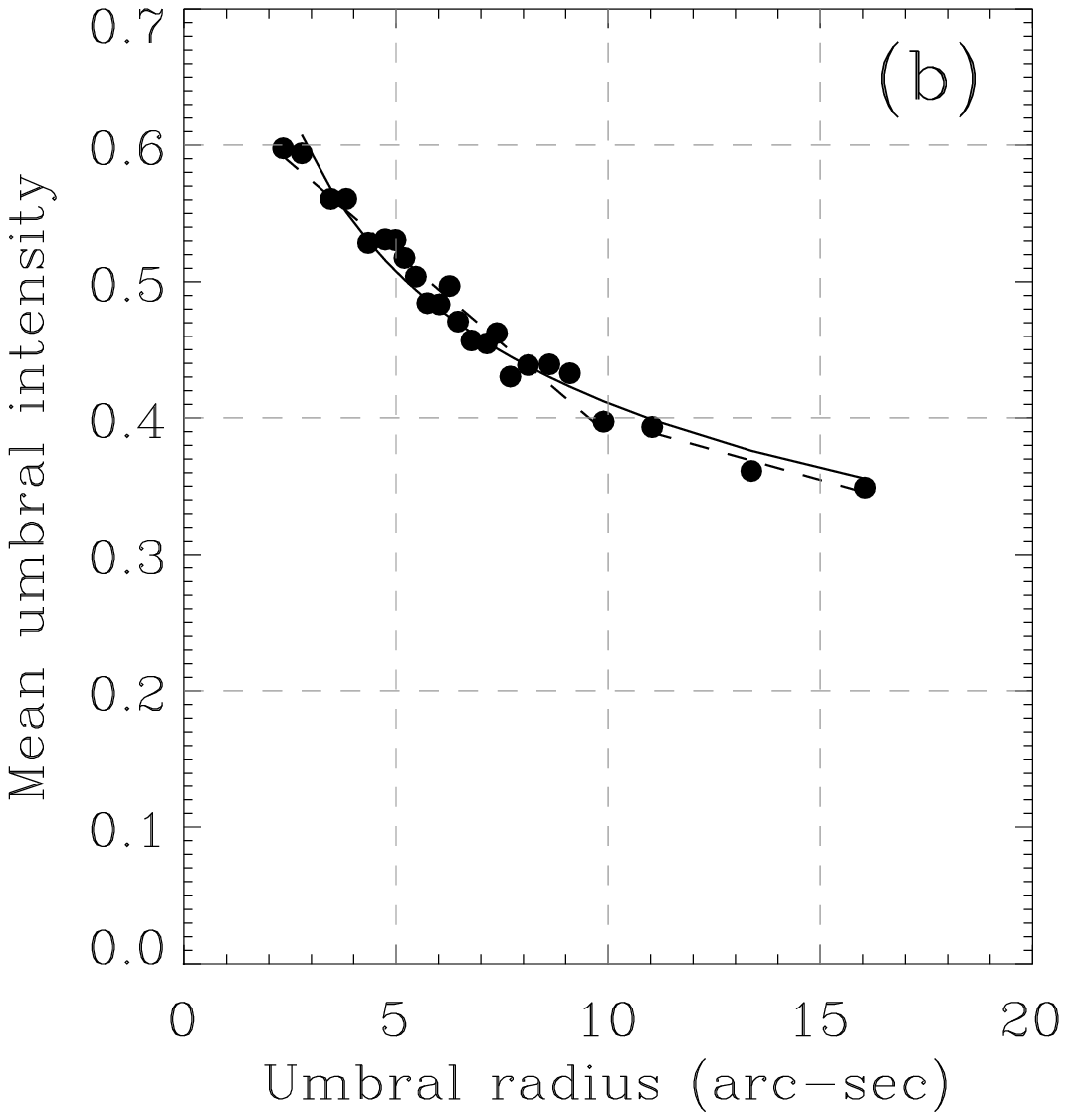}} 
\caption{Power law fit (solid line) and double linear fit (dash lines) to the {\bf (a)} 
umbral core intensity and, {\bf (b)} mean umbral intensity. 
Here the filled circles represent bins of 10 spots each.} 
\end{figure}
\begin{figure}
\resizebox{\hsize}{!}{\includegraphics{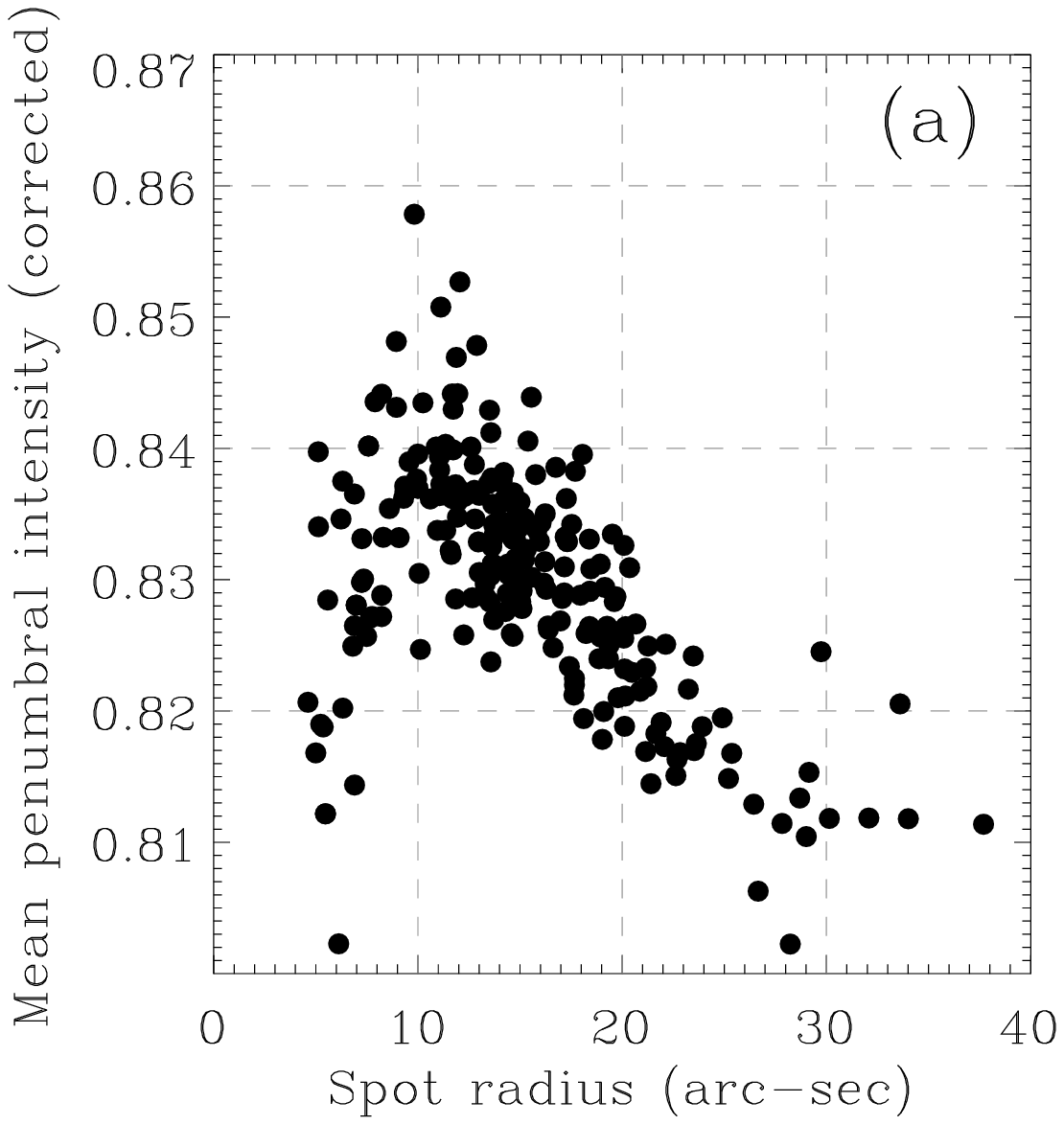}\includegraphics{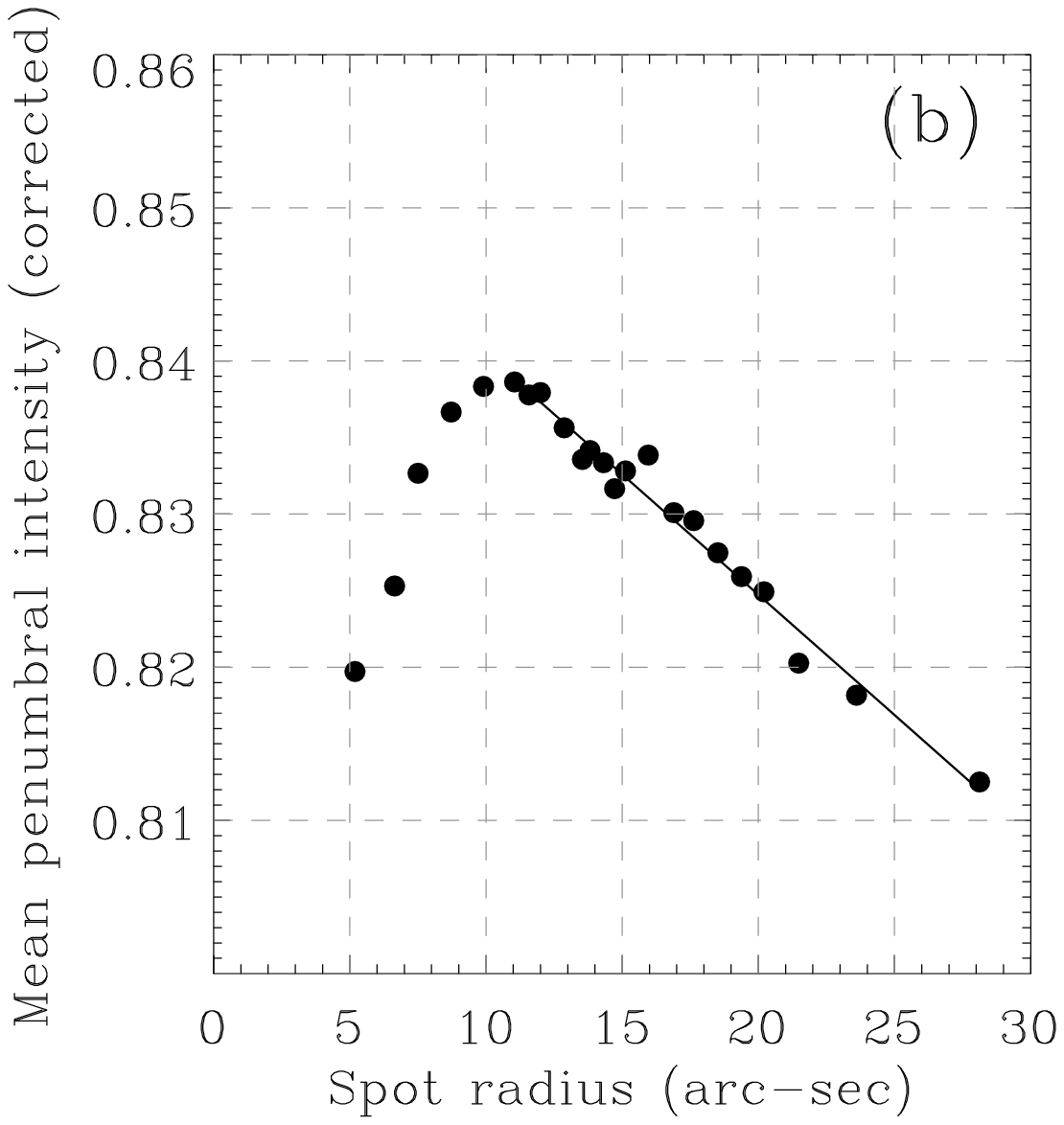}}
\caption{{\bf (a)} Mean penumbral intensity versus spot radius, corrected for stray light. 
{\bf (b)} Linear fit to the binned mean penumbral intensity.}
\end{figure}
\begin{table*}
\caption{Fit parameters for radius-brightness relation}
\begin{tabular}{lcccccc} 
\hline 
\hline
 Dependence on umbral radius of,& Umbral radius	& Constant & Exponent & Gradient & $\sigma $	& $\chi^{2}$ \\
\hline
\bf{Power law fit} 	&		&		&	        &	       &  & \\
umbral core intensity 	&(all)  	& 1.8598	& $-1.0679$   &	&0.063       &$3.5 \times 10^{-3}$\\
umbral mean intensity 	&(all) 		& 0.8297        & $-0.3052$   & &0.013	      &$1.5 \times 10^{-4}$ \\ 
\bf {Double linear fit} 	&		&		&		&		& \\
umbral core  intensity	        &$<$10$''$ 	&0.6515  &	&$-0.0552$ 	&0.0029 	&$6.7 \times 10^{-4}$\\
			        &$>$10$''$	&0.2299  &	&$-0.0094$ 	&0.0027  	&$4.7 \times 10^{-5}$\\
umbral mean intensity	        &$<$10$''$ 	&0.6536  &	&$-0.0266$ 	&0.0013  	&$1.2 \times 10^{-4}$ \\
			        &$>$10$''$	&0.4858  &	&$-0.0087$ 	&0.0026 	&$4.3 \times 10^{-5}$\\ 
\hline 
Dependence on spot radius of, 	&		&		&		&		&\\
\hline
\bf {Linear fit} 		&		&		&		&		&\\
penumbral mean intensity 	&$>$10$''$	&0.8561  &	&$-0.0016$   	&0.0001 	&$1.1\times 10^{-6}$\\ 
\hline 
\hline
\end{tabular}
\end{table*}

A steeper decrease 
is found for  spots with smaller umbral radius, while for the bigger 
spots a more gentle decrease in umbral core intensity with radius is observed 
(dictated by the fact that umbral intensity has to be positive).  
This plot emphasises the need to take into account the dependence of the 
umbral brightness on the size of the spot  when looking for solar cycle variations.

The mean umbral intensity, plotted in Fig. 11, also shows a similar decrease 
with increase in umbral radius. The difference between the umbral core 
and mean intensities is smallest for the smallest umbrae and increases with umbral size. 

In order to obtain a relation between the umbral core and mean intensities with umbral radius,
we carried out two different fits to the respective corrected umbral intensities after binning 
together points with similar sunspot radius, such that each bin contains 10 samples. The dashed lines 
in Figs. 12(a) and (b) show the double linear 
fits to the  umbral core and mean intensities, respectively. The individual linear fits are made 
to  spots with umbral radii less than 10\arcsec\/ and to those with radii above this limit, respectively.
The fit parameters along with errors and normalised $\chi^{2}$ values are included in Table 2. 
The solid lines in these figures show the power law fit. The power law fit to the umbral 
core intensity seems to be a comparatively  poor approximation, while the mean umbral brightness 
seems to obey the power law (i.e. it gives a very low $\chi^2$ for half the number of free
parameters as the double linear fit). The parameters for the power law fit are also included 
in Table 2. Figure 12 again demonstrates that the difference between the core and mean intensity 
increases rapidly with umbral radius. It is equally clear that since the umbral core intensity 
varies by a factor of nearly 6, the mean umbral intensity by a factor of nearly 2 between the smallest 
and the largest umbrae, employing a single value for  umbral brightness of all spots is a very poor 
approximation. Such an approximation is often made e.g. for the reconstruction of solar irradiance 
(cf. Unruh et al. \cite{unruh}, Krivova et al. \cite{krivova}).

\subsection{Penumbral mean intensity and spot radius}
Figure 13(a) shows the relation between mean penumbral intensity  and spot radius. 
An approximate linear relationship is evident for the spots with outer penumbral radius between 
10\arcsec\/ and 30\arcsec. The outer penumbral radius is the equivalent radius of the whole sunspot 
(including the umbra). The large scatter in mean penumbral intensities for spot 
sizes below 10\arcsec\/ might  result from the insufficient resolution of the 
full disk images or may be due to the fact that the parameters for distinguishing 
between umbra and penumbra are possibly not appropriate for small spots. Note the 
order of magnitude smaller range of variation of penumbral contrast than of 
umbral contrast. Figure 13(b) shows the linear fit to the mean penumbral brightness, 
after binning 10 adjacent spots (taking only spots with radius greater than 10$''$). 
The fit parameters are listed in Table 2.        
\begin{figure*}
\sidecaption
\includegraphics[width=12cm]{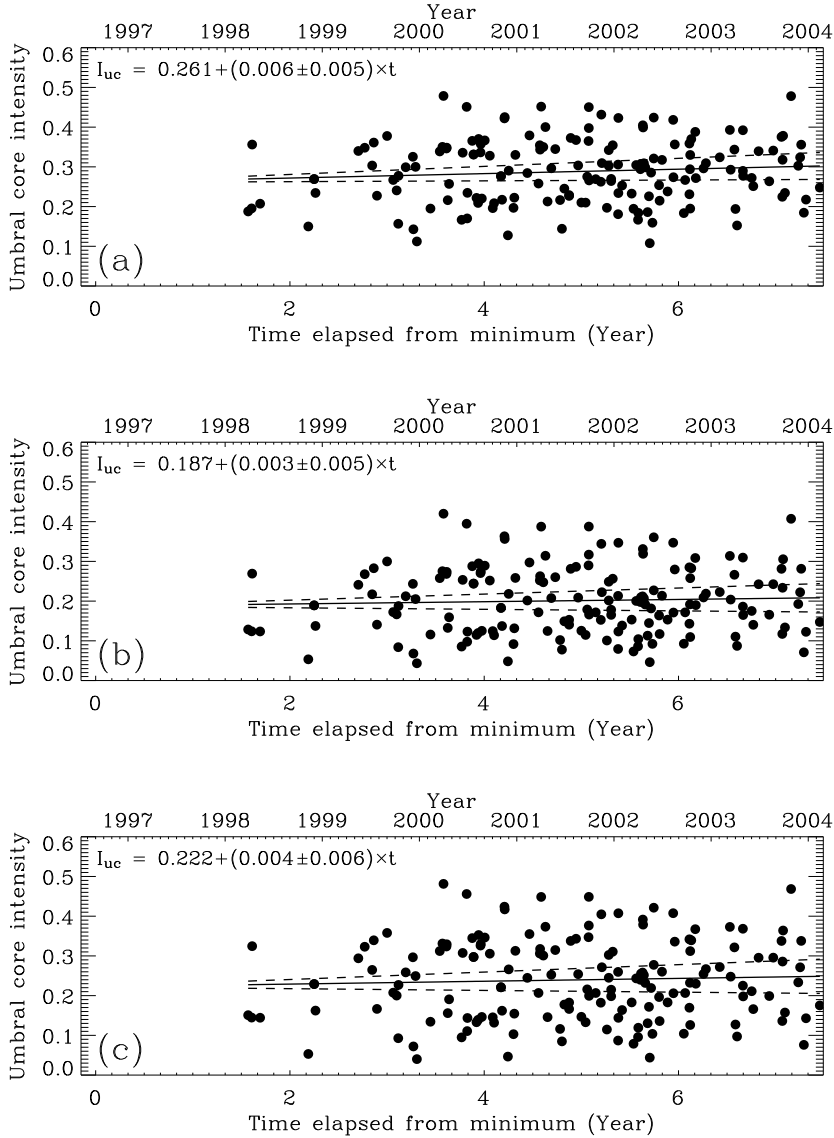}
\caption{Umbral core intensity versus solar cycle, {\bf (a)} observed, {\bf (b)} corrected for 
stray light, and {\bf (c)} corrected for stray light and the influence of Ni~{\scriptsize I} \/line. 
The intensities are plotted for all spots with umbral radii between 
5 arc-sec and 15 arc-sec. The solid line shows the linear regression 
and the dashed lines represent the $\pm1 \sigma$ deviation, due to the uncertainty in the regression 
gradient. The best linear fit is given in the upper left corner.}
\end{figure*}

\section{Solar cycle dependence of the brightness}
In Fig. 14 we plot the sunspot umbral core intensity versus time elapsed since the solar cycle 
minimum. September 1996 is taken as the minimum month, the upper axis shows the corresponding 
year. This plot includes spots with umbral radius between 5\arcsec\/ and 15\arcsec\/ only, in order to be 
consistent with the work of Albregtsen \& Maltby (\cite {albregtsen1}). Figure 14(a) shows the observed 
intensity, (b) the stray light corrected intensity, while  Fig. 14(c) shows the intensity corrected for both 
stray light and the influence of the Ni~{\scriptsize I} \/line on the continuum measurements. In all the figures 
the trend remains the same. As the umbral radius and core brightness are related, the scatter in one quantity 
also reflects the scatter in the other.

The solid line shows the linear regression to the brightness, whereas the dotted 
lines indicate the 1$\sigma$ error in the gradient. A feeble trend of increasing umbral 
brightness towards the later phase of the solar cycle is observed. This increase is 
well within the  1$\sigma$ error bars and statistically insignificant. Table 3 lists the fit 
parameters, including the errors and normalised $\chi^{2}$ values. All the fit parameters 
are listed for the corrected intensity values. Also, in all the remaining figures we plot  
the corrected intensities alone. 
\begin{figure*}
\includegraphics[width=12cm]{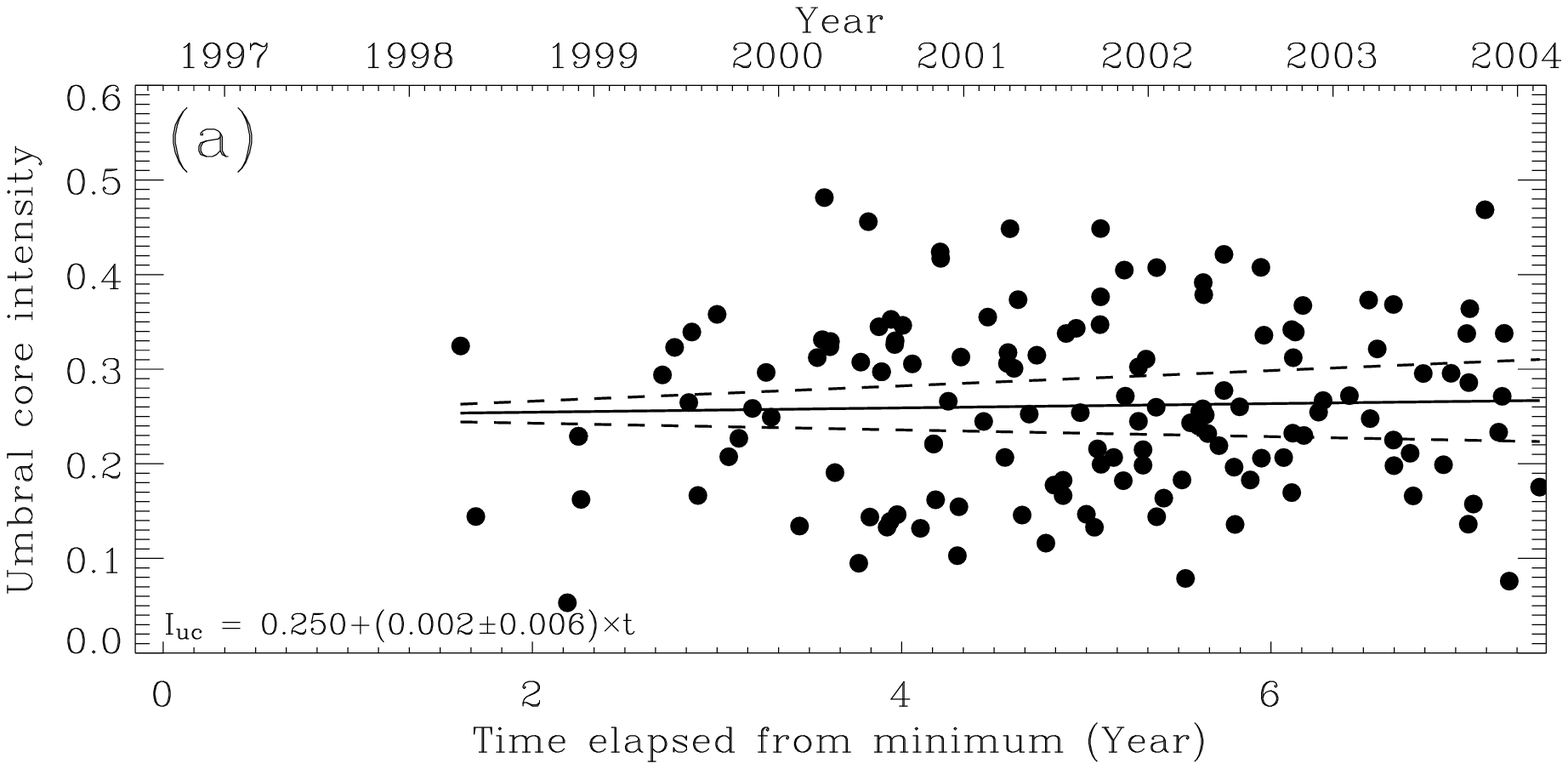}\\
\sidecaption
\includegraphics[width=12cm]{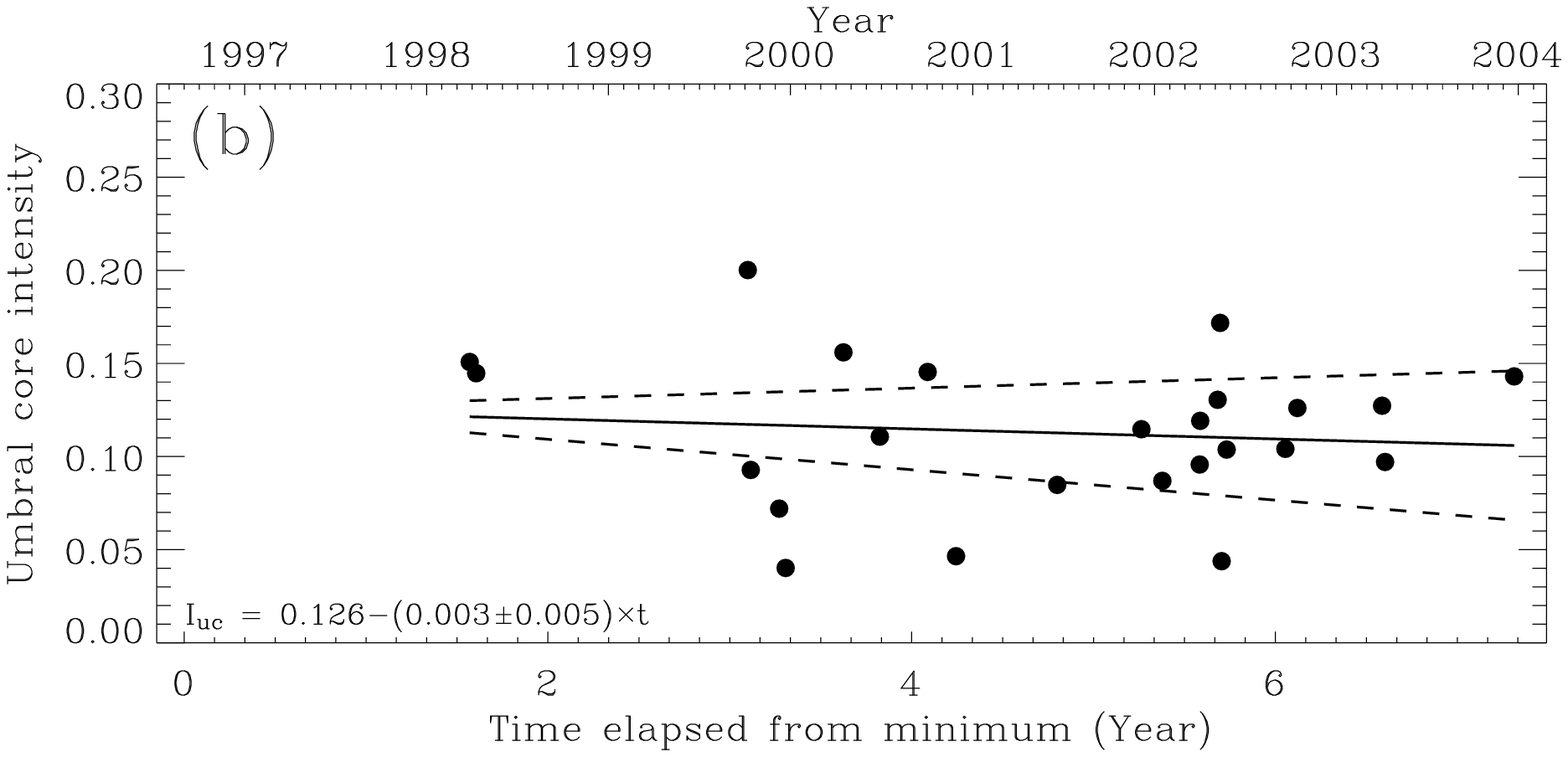}
\caption{Umbral core intensity versus solar cycle plotted for {\bf (a)} umbral radii 
ranging from 5 to 10 arc-sec and, {\bf (b)} from 10 to 15 arc-sec.  
The solid lines show the regression fits and the dashed lines the $\pm1 \sigma$ deviations 
due to the uncertainty in the regression gradient.}
\end{figure*}
\begin{table*}\caption{Fit parameters for solar cycle dependence}
\begin{tabular}{lcccccc}\hline \hline
Solar cycle dependence of, 	&Umbral radius		&Constant 	& Gradient	& $\sigma $ 	& Gradient$/ \sigma $	& $\chi^{2} $ \\ \hline
umbral core intensity 		& (all)			&0.33605	&$-0.00785$	&0.00649	&$-1.2095$		&0.01901\\
				&5$''$ - 15$''$ 	&0.22188 	&$+0.00358$   	&0.00577 	&$+0.6205$		&0.01015\\ 
				&5$''$ - 10$''$ 	&0.25001	&$+0.00227$ 	&0.00582 	&$+0.3900$		&0.00841\\
				&10$''$ - 15$''$	&0.12560 	&$-0.00269$ 	&0.00547 	&$-0.4918$		&0.00158\\
				&(all, northern hemisphere)&0.33957	&$-0.01071$	&0.00622	&$-1.7218$		&0.00085\\
				&(all, southern hemisphere)&0.33062	&$-0.00310$	&0.00874	&$-0.3547$		&0.00164\\\hline
umbral mean intensity 		&5$''$ - 15$''$ 	&0.44346 	&$+0.00206$   	&0.00312 	&$+0.6603$		&0.00296\\ 
				&5$''$ - 10$''$ 	&0.45782	&$+0.00168$ 	&0.00294 	&$+0.5714$		&0.00215\\
				&10$''$ - 15$''$	&0.39609 	&$-0.00333$ 	&0.00513 	&$-0.6491$		&0.00139\\\hline
penumbral mean intensity 	&5$''$ - 15$''$ 	&0.82974	&$-0.00009$   	&0.00049 	&$-0.1837$		&0.00007\\ 
				&5$''$ - 10$''$ 	&0.83324	&$-0.00034$ 	&0.00042 	&$-0.8095$		&0.00004\\
				&10$''$ - 15$''$	&0.81723	&$-0.00024$ 	&0.00074 	&$-0.3243$		&0.00003\\\hline
umbral radius 		        & (all)			&5.79088	&$+0.19029$	&0.13549	&$+1.4045$		&8.29028\\
				&5$''$ - 15$''$ 	&8.06675	&$-0.07457$ 	&0.12525 	&$-0.5954$		&4.77502\\
				&5$''$ - 10$''$	   	&7.06682	&$-0.01697$ 	&0.08311 	&$-0.2042$		&1.71632\\
				&10$''$ - 15$''$ 	&11.5536	&$+0.06823$ 	&0.20443 	&$+0.3338$		&2.20379\\
				&(all, northern hemisphere)&5.75164	&$+0.21357$	&0.11716	&$+1.8229$		&0.30162\\
				&(all, southern hemisphere)&5.94171	&$+0.12221$	&0.19719	&$+0.6198$		&0.83724\\
\hline
\hline
\end{tabular}
\end{table*}
\begin{figure*}
\includegraphics[width=12cm]{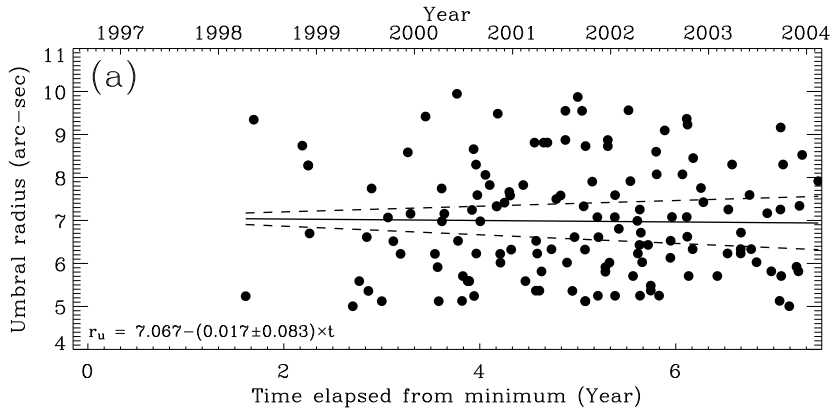}\\
\sidecaption
\includegraphics[width=12cm]{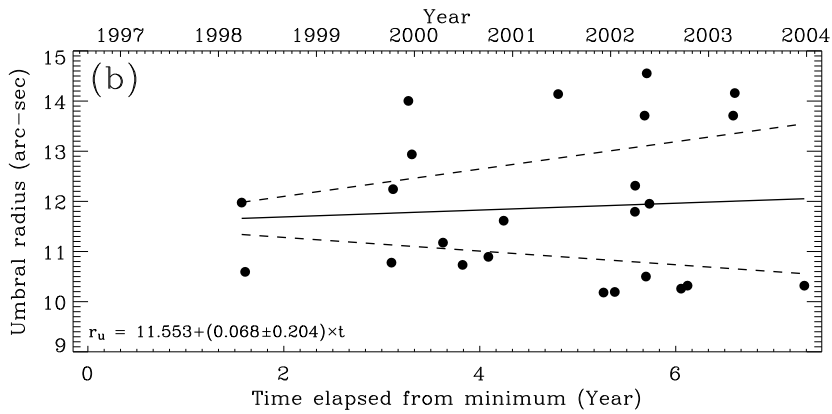}
\caption{Solar cycle dependence of umbral radius for {\bf (a)} sunspots with radii between 
5 and 10 arc-sec and {\bf (b)} with radii between 10 and 15 arc-sec.}
\end{figure*}  
\begin{figure*}
\sidecaption
\includegraphics[width=12cm]{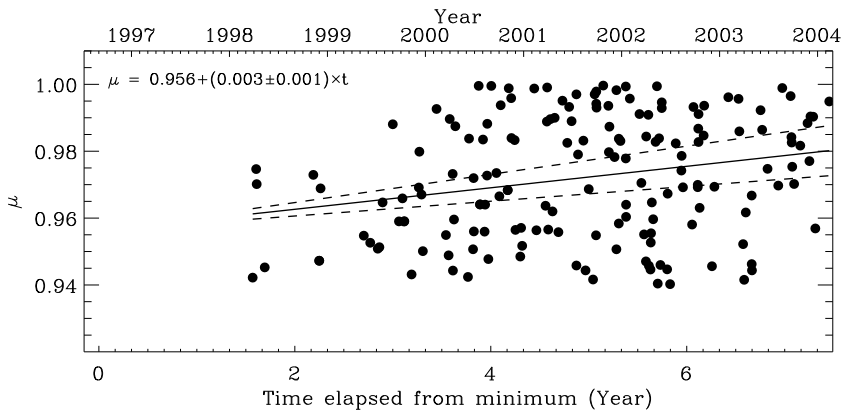}
\caption{$\mu$ versus time since activity minimum for all spots with umbral radii between 
5 arc-sec and 15 arc-sec. The solid line shows the linear regression 
and the dashed lines the $\pm1 \sigma$ deviation.}
\end{figure*}
\begin{figure}
\resizebox{\hsize}{!}{\includegraphics{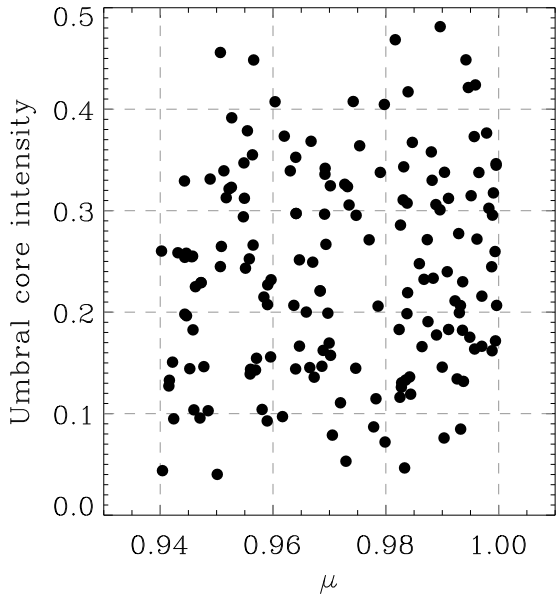}}
\caption{Umbral core intensity versus $\mu$ for all spots with umbral radii between 
5 arc-sec and 15 arc-sec. 
}
\end{figure} 
\begin{figure*}
\sidecaption
\includegraphics[width=12cm]{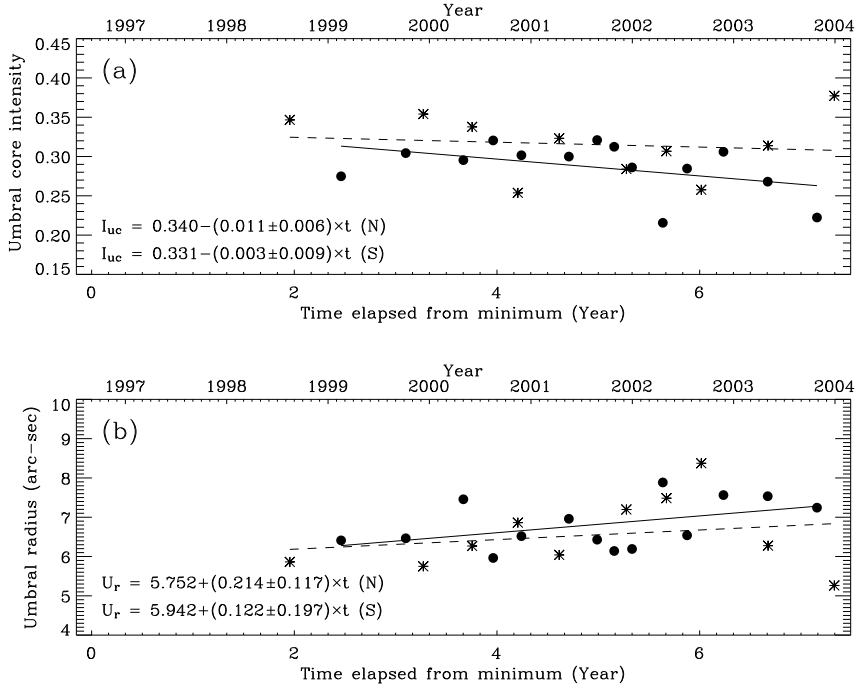}
\caption{Umbral core intensity {\bf (a)} and umbral radius {\bf(b)} versus time since solar cycle minimum 
for Northern (filled circles) and Southern (asterisks) hemispheres, for all observed sunspots. 
Each plotted symbol represents an average over 10 sunspots.  The solid  and dashed lines show the linear regression 
fits for Northern and Southern hemispheres, respectively. The best linear fits are given in the lower 
left corner.}
\end{figure*}
\begin{figure}
\resizebox{\hsize}{!}{\includegraphics{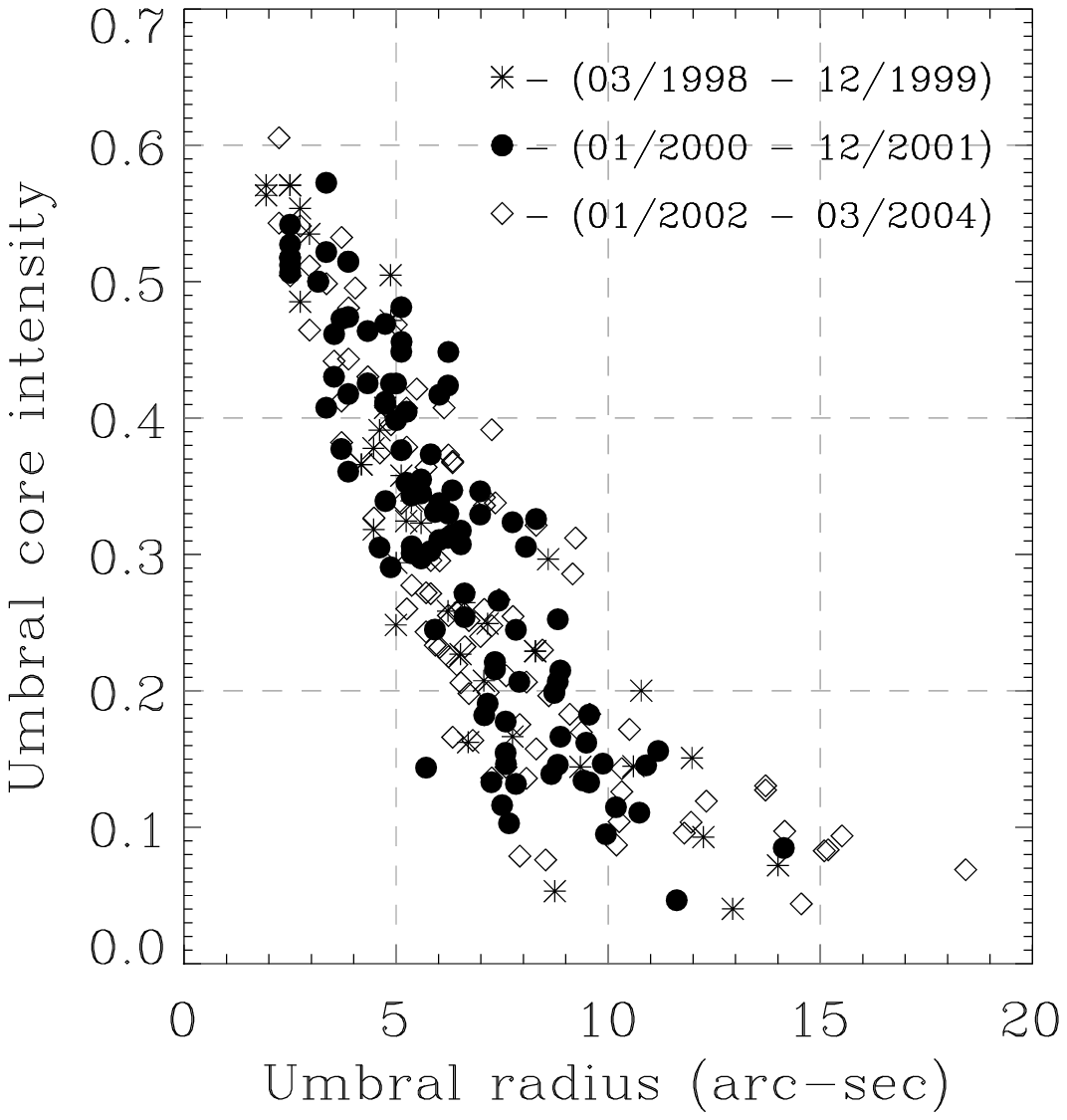}}
\caption{Umbral core intensity versus umbral radius. Different symbols represent spots observed 
in ascending, maximum and descending phases of solar cycle. The dates indicate the period of observation.}
\end{figure}  
In Figs. 15(a) and (b) we display the umbral 
core intensity for two different umbral size ranges, i.e. for spots with umbral 
radii  in the range 5\arcsec\/ - 10\arcsec\/ and in 10\arcsec\/ - 15\arcsec, respectively. 
The trend seen here is opposite for small and large spots, but is insignificant 
in both  cases (see column Gradient/$\sigma$ in Table 3). If we plot all the sunspots, 
irrespective of radius, the results are not significantly different.
 
In Figs. 16(a) and (b) we plot the dependence on the time elapsed from 
activity minimum, of the analysed  sunspot umbral radius, separately for the 
small (5\arcsec\/ - 10\arcsec) and the large (10\arcsec\/ - 15\arcsec) spots, respectively. 
The linear regression  are overplotted. 
The mean umbral radius of the analysed spots   
between 5\arcsec\/ - 10\arcsec\/ slightly decreases with time, whereas spots 
with radius between 10\arcsec\/ - 15\arcsec\/ show the opposite trend. Similarly in 
Fig. 1 the umbral radii of all studied spots are plotted. The regression parameters
for the full sample of spots (5\arcsec\/ - 10\arcsec) is given in Table 3. Although none of these
trends is statistically significant, they are opposite to the trends (also not 
statistically significant) shown by the umbral brightness of these spots. This 
is completely consistent with the dependence of brightness on umbral radius shown 
in Figs. 10 - 12.

Another bias can be introduced by the fact that the sunspot latitude systematically 
decrease over a solar cycle. We have to a certain extent reduced this effect by 
considering only sunspots at $\mu > 0.94$. Figure 17 shows the average $\mu$ of the 
analysed sunspots, as expected, this displays an increase over the cycle.  
In order to judge whether this introduces a bias into the cycle phase dependence of sunspot 
brightness we plot in Fig. 18 umbral core brightness (i.e. contrast to local quiet Sun) 
versus $\mu$ (for $0.94 < \mu < 1$). We did not find any significant variation in the umbral 
core brightness with $\mu$.

In order to check for an asymmetry in umbral brightness between the northern and southern 
hemispheres as reported by Norton \& Gilman (\cite{norton}), in Fig. 19  we plot umbral core 
brightness separately for northern and southern hemispheres. In this plot we used corrected 
intensities for all the observed spots. The intensities are binned and each bin contains 10 samples. 
The linear regression fit provides a slightly higher gradient in the northern hemisphere. 
But this can be well explained by the increase in umbral 
radius of the spots with the cycle phase (Fig. 19(b)). It should be noted that Norton \& Gilman observed a significant 
umbral brightness difference between the northern and southern hemisphere during the onset of cycle 23. 
Due to the restriction of $0.94 < \mu $ in our selection criteria, we have analysed only few spots  during 
this period and hence cannot comment on that result.

In Table 3 we also list the parameters of the linear regressions to mean umbral and 
penumbral intensities versus time. None of the gradients is significant at even the 
1$\sigma$ level. Also, the signs of the gradients of all umbral core and mean intensity 
samples are opposite to those of the umbral radius of the corresponding sample, 
suggesting that even any small gradient in the umbral brightness is due to a small 
bias in the umbral size with time. Hence we find no evidence at all for a change 
in sunspot brightness over the solar cycle. 

In order to test whether the dependence of umbral core brightness on umbral radius in 
Sect. 4.1 is itself dependent on solar cycle phase we plot in Fig. 20 the umbral core brightness 
versus radius but now for three different phases of the cycle. Asterisks, filled circles and 
diamond symbols represent the spots observed in ascending, maximum and descending phases of solar 
cycle, respectively.  As can be clearly seen there is no difference between the different phases. 
This demonstrates that there is no cross-talk between cycle phase dependence and umbral radius 
dependence of umbral brightness.    
    
\section{Discussion}
With  a large sample of sunspots, we have tested 
if the umbral core brightness, the umbral average brightness, or the penumbral 
brightness depend on solar cycle phase. In addition to this, we 
studied the  dependence of brightness on sunspot size.

Earlier continuum observations suggested that large sunspots are 
darker than smaller sunspots (Bray \& Loughhead \cite{bray}). But most 
of such observations were barely corrected for stray light (Zwaan \cite{zwaan}).  Subsequent 
observations which were corrected for stray light showed no significant 
dependence of umbral core brightness on spot size (Albregtsen \& Maltby 
\cite{albregtsen2}). More recent observations however, reveal that even after stray-light 
correction a size dependence remains. Thus, Kopp \& Rabin (\cite{kopp}) 
present observations at 1.56 ${\rm {\mu m}}$ that show clear evidence for the size dependence 
of umbral brightness. Also,  results from two sunspots observed at the same 
wavelength combined with the Kopp \& Rabin data confirm and strengthen the 
linear dependence of brightness on sunspot umbral size (Solanki \cite{solanki1997}, 
Solanki et al. \cite{solanki1}, Kopp \& Rabin \cite{kopp}, R$\mathrm{\ddot{u}}$edi et al. \cite{ruedi}).

Mart\' \i nez Pillet \& V$\acute{\rm a}$zquez (\cite{martinez}) confirmed the Kopp \& Rabin  result 
based on the analysis of 7 sunspots. Collados et al. (\cite{collados}) found from the inversion 
of Stokes profiles obtained in 3 sunspot umbrae that small umbrae are distinctly 
hotter than large umbrae. One disadvantage of all these studies is that each is restricted to a
relatively small number of sunspots. Another is that each sunspot was observed under 
different seeing conditions, so that the level of stray light varied in an unsystematic
manner. Both shortcomings are addressed in the present paper. 

In our analysis we found a clear dependence of umbral core brightness on 
umbral size. Since we correct for the very slowly varying stray light and the MDI specific 
problem of cross-talk of the spectral line into the continuum, as described 
in Sect. 3 and the appendix, our results are basically free from stray light contamination. 
This is particularly true for umbral core intensities as we could show using the Mercury transit data.
For mean intensities some residual remains (see Appendix A).
Also, we have a relatively large sample of sunspots  to support our results. 
We carried out and compared two different fits to the umbral brightness-size 
dependence for the analysed spots, a double linear fit for two different umbral 
size ranges and a power law fit to the entire data set. A similar analysis of 
umbral brightness and diameter carried out at a wavelength of 1.56 $\rm \mu m$  
(Solanki \cite{solanki1997}) shows a smaller gradient than what we obtained in our 
work. Umbral brightness for 9 spots, with umbral diameter ranging from 5\arcsec\/ 
to 35\arcsec\/  were included in the above study. From a linear fit, 
a gradient of around $-0.012/\arcsec$ is obtained for the umbral brightness, 
whereas in our case a much higher gradient of $-0.04/\arcsec$ is found. 
This is not surprising since the brightness at 1.56 $\rm \mu m$  reacts much more weakly
to a given temperature change than at 677 $\rm {nm}$.
  
Albregtsen \& Maltby (\cite{albregtsen1}) reported a variation of umbral 
core brightness with solar cycle by analysing  13 sunspots. The results 
were mainly presented for the observations done at a wavelength of 1.67$\rm {\mu m}$. 
They found an increase in umbral core intensity by as much as 0.15 from early to 
late phase of the solar cycle. They also found no dependence of umbral brightness 
on other sunspot parameters, such as size and type of the spot. In a later 
paper Maltby et al. (\cite{maltby}) detailed this variation for a range of wavelengths 
starting from 0.38 - 2.35 $\rm {\mu m}$. The nearest wavelength to our observations 
(i.e. 0.669 $\rm {\mu m}$) shows a variation of around 0.072 in umbral core intensity from 
early to late phase, this corresponds to 0.0065 umbral intensity variation per year. 
Based on these findings they presented three different umbral 
core model atmospheres  for sunspots present in the early, mid and late phase of 
the solar cycle. In a recent study carried out using MDI data Norton \& Gilman 
(\cite {norton}) find a relatively smooth decrease in the umbral brightness from activity minimum 
to maximum for Northern hemisphere and no distinct trend for the southern hemisphere.   
The decrease in umbral brightness with solar cycle they found is opposite to the results 
of Maltby et al. (\cite{maltby}).      

In our analysis we found a very feeble, statistically  insignificant  dependence of 
umbral brightness on solar cycle (i.e. any change remains well within the error bars).
The linear fit to umbral core brightness is given by the following 
equation,
\begin{eqnarray}
I_{uc}=0.222+(0.004\pm 0.006)\times t
\end{eqnarray}
where $t$ is the time elapsed from the minimum in units of years. 
All sunspots within 5\arcsec\/ - 15\arcsec\/ umbral radius are included in the regression. 
It is striking that the 1$\sigma$ uncertainty in the gradient obtained in our 
analysis is approximately equal  to the trend found by Maltby et al. (\cite{maltby}).
Therefore, either the MDI wavelength is less suitable than the 1.56 $\rm \mu m$ wavelength 
band employed by Maltby and co-workers, or there is a selection bias affecting their results. 
Indeed, the change in umbral core intensity over the solar cycle reported by 
Maltby et al. (\cite{maltby}), 0.072, is only $1/3$ the umbral core intensity difference 
between spots with umbral radii of 5\arcsec\/ and 15\arcsec\/ found here and is 0.6 times the 
intensity difference between such spots at 1.56 $\rm \mu m$. Consequently, selection biases, 
which often afflict small samples, can introduce an artificial 
trend of the correct magnitude over an activity cycle. 

In order to reduce the effect of size dependence on the above relation, we grouped the sunspots 
into two umbral radii bins.  The linear regressions  to these groups are given by the following 
equations,
\begin{eqnarray}
I_{uc}= \{
\begin{array}{ll}
0.250+(0.002\pm0.006)\times t &{\rm for} ~~5'' - 10''\\
0.126-(0.003\pm0.005)\times t &{\rm for} ~~10''- 15''  
\end{array}
\end{eqnarray}
For the group of spots with umbral radii between  5\arcsec\/ - 10\arcsec\/ the linear regression 
fit gives an increase in umbral brightness with increasing phase of the solar cycle, 
but again the change is within the error bars. For the spots with umbral radii 
10\arcsec\/ - 15\arcsec\/ the opposite trend is found. We compared these results with the 
variation of the sizes of the 
analysed sunspots over the solar cycle. It turns out that for all three samples 
(5\arcsec\/- 10\arcsec\/, 10\arcsec\/-15\arcsec\/, 5\arcsec\/-15\arcsec\/) the average radii 
show the opposite trend to the intensity. This suggests that at least part of any trend in 
brightness over the solar cycle is due to a corresponding (opposite) trend in umbral radii.

From these results it is evident that the 
dependence of the brightness of the spot on size is an important parameter to 
be considered when the umbral brightness variation with solar cycle is studied. We believe that 
without showing the time dependence of the average area or size of the sunspot umbrae in the employed 
samples, any studies of sunspot brightness (or even field strength) evolution over time are of limited 
value. Although there is no systematic variation in the relative distribution of 
umbral areas with solar cycle (Bogdan et al. \cite {bogdan}) in a smaller sunspot sample
a trend may be introduced by limited statistics. It would therefore be of great value to determine
the areas of the umbrae studied by Penn \& Livingston (\cite{penn}), who find a steady decrease of the 
maximum umbral field strength over the last 7 years, since the field strength is related to brightness 
(Maltby \cite{maltby2}, Kopp \& Rabin \cite{kopp}, Mart\' \i nez Pillet \& V$\acute{\rm a}$zquez \cite{martinez}, 
Solanki et al. \cite{sol1993}, Mathew et al. \cite{mathew2004}, Livingston \cite{livingston}), 
which depends on size (Sect. 4.1). Also, our results imply that models of the umbral 
atmosphere (e.g. Avrett \cite{avrett}, Maltby et al. \cite{maltby}, Caccin et al. \cite{caccin}, 
Severino et al. \cite{severino}, Fontenla et al. \cite{fontenla}) should always indicate the spot 
size to which they refer.

Chapman et al. (\cite{chapman})
reported photometric observations of sunspot groups, that show considerable variation in their 
mean contrast. This could be due to umbra/penumbra area ratio change, or due to intrinsic 
brightness change. Our results suggest that, at least in part the later is the reason. This has 
implications for irradiance reconstructions, particularly those using separate atmospheric 
components for umbrae and penumbrae (e.g. Fligge et al. \cite{fligge}, Krivova et al. \cite{krivova}).

The result is also of importance for the physics of sunspots, since an explanation 
must be found why a smaller heat flux is transported through the umbrae of larger sunspots. 
E. g. in Parker's (\cite{parker}) spaghetti model it would imply that either the filamentation 
of the subsurface field  is less efficient under larger umbrae, or that the energy flux transported 
between the filaments is smaller for larger filaments. One factor which probably plays a role is that
darker sunspots have higher field strengths (Maltby \cite{maltby2}, Kopp \& Rabin  \cite{kopp}, 
Livingston \cite{livingston}) which are more efficient at blocking magnetoconvection. 
This also implies that larger sunspots have stronger fields.      
 
The only clear dependence we have found in our sample of sunspots is between 
umbral intensity and radius. However, even this relationship shows considerable 
scatter, whose origin is not clear. We list some possibilities below:
\begin{enumerate}
\item dependence of the scattered light correction on the shape of the umbra 
(e.g. very elongated versus circular);
\item dependence of intrinsic brightness of umbrae on shape and complexity;
\item dependence of brightness on age;
\item (small) dependence of brightness on phase of the solar cycle;
\item CLV of umbral contrast (which is small according to Fig. 18).
\end{enumerate}
Separating clearly between these possibilities is beyond the scope of 
the current paper. Our analysis is restricted to regular sunspots with single umbrae. 
Complex spots may in principle show a different behaviour. 
\section{Concluding remarks}
In this paper we present the analysis of MDI continuum sunspot images aimed at 
detecting umbral core brightness variation with solar cycle. We analysed a total of 
234 sunspots of which 164 sunspots have an umbral radius lying between 5\arcsec\/ - 15\arcsec. 
Careful corrections for stray light and the Zeeman splitting of the nearby 
Ni~{\scriptsize I} \/line on measured continuum intensities have been made.
We derive the following conclusions from our analysis. 
\begin{itemize}
\item The umbral core and mean brightness decreases substantially  with increasing  umbral radius. 
\item The mean penumbral intensity is also reduced with increased spot size, but by a small amount.
\item No significant variation in umbral core, umbral mean and mean penumbral intensities is 
found with solar cycle.
\item The insignificant variation with solar cycle of the umbral intensity could be at least 
partly  be explained by the dependence of the analysed spot size on 
solar cycle.
\end{itemize}
\begin{acknowledgements}
We wish to express our thanks to SOHO/MDI team for providing the full-disk continuum 
intensity images. Thanks are also due to Dr. Andreas Lagg for providing the updated 
code for computing the line profiles and the referee Aimee Norton for her useful suggestions. 
This work was partly supported by the Deutsche Forschungsgemeinschaft, DFG project number 
SO~711/1-1. Funding by the Spanish National Space Program (PNE) under project ESP2003-07735 
is gratefully acknowledged.  
\end{acknowledgements}
\appendix
\section{Stray-light evaluation and image restoration of MDI data}

The point spread function used to fit the limb profiles follows Mart\'\i nez
Pillet (\cite{valentin}) and it includes three Gaussians
and one Lorentzian component:
\begin{figure*}
\includegraphics[width=17cm,height=10.0cm]{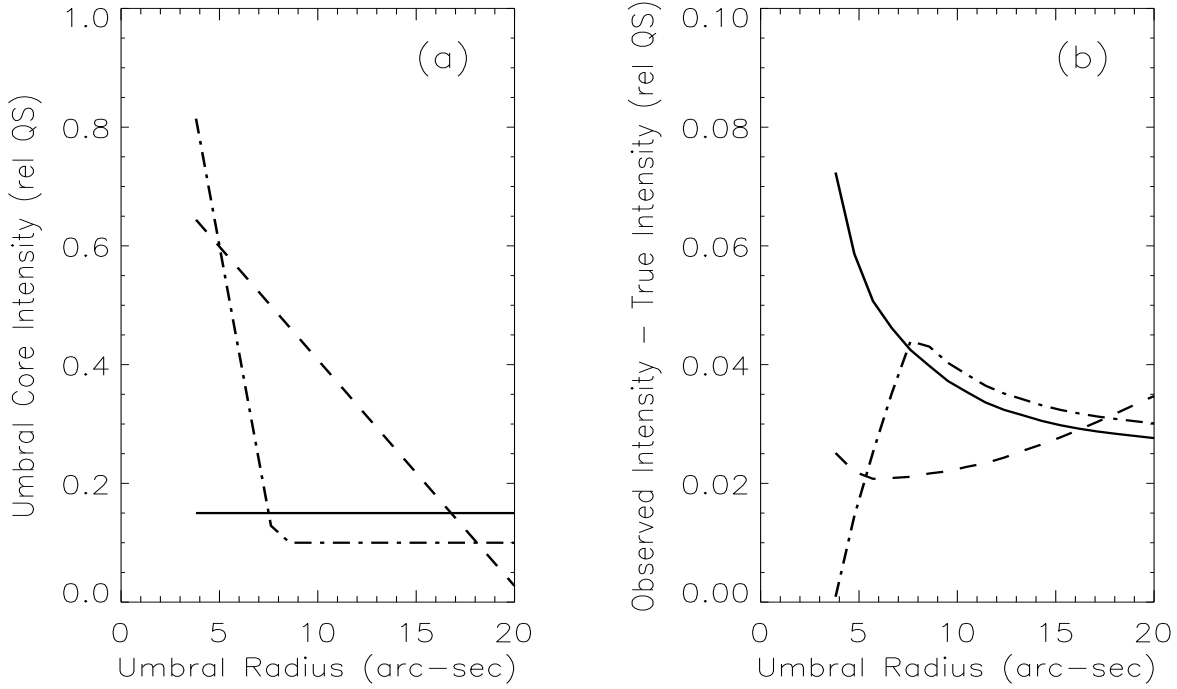}
\caption{Dependence between umbral intensity and size {\bf (a)} used to compute the 
contamination levels (observed intensity minus true intensity) shown in {\bf (b)}.
Solid lines correspond to a constant umbral intensity for all umbral sizes. Dashed lines
correspond to a linear decrease of umbral intensity with umbral radius. Dashed-dot 
lines represent a combination of the two previous cases. The spread functions
used are those derived from the Mercury transit observations.}
\end{figure*}

\begin{equation}
\psi(r)=\sum_{i=1}^{3} m_i a_i exp(-{{r^2}\over{b_i^2}})+M{{A}\over{B^2+r^2}} {\bf ,}
\end{equation}
where $r$ is the radial distance from disk centre, $m_i$ and $M$ the weights of the different
components (that add to 1), $b_i$ the corresponding Gaussian parameters, $B$ the Lorentzian parameter
and $a_i$ and $A$ the normalisation constants (see 
Mart\'\i nez Pillet \cite{valentin}, equations 23 and 24).
The theoretical CLV used for the (two-dimensional) convolution with the spread function is 
a simple polynomial:
\begin{equation}
{{I(r)}\over{I_0}}=\sum_{i=1}^{6} c_i \mu^i(i-1) {\bf  ,} 
\end{equation}
with $\mu$ the cosine of the heliocentric angle. Note that all
polynomial coefficients, $c_i$, add to one.
The azimuthally averaged MDI continuum intensity profiles are fitted with a standard
non-linear least-squares algorithm. The free parameters are the Gaussian and Lorentzian parameters
($b_i$ and $B$), the three Gaussian weighting parameters ($m_i$) and five of the
CLV coefficients ($c_i$). In total there are 12 free parameters. While we originally left fixed the
CLV coefficients to values found in the literature, it was clear from the residuals observed inside 
the solar disk that they should be left also as free parameters (note also that the continuum of MDI
does not correspond to a clean continuum, as explained in the text).  

In Table A.1, we provide the spread function fitted parameters for four frames observed during the Mercury 
transit on the 7th of May, 2003.  The corresponding limb profiles were all very similar and showed 
differences at the few $10^{-4}$ level (of the disk centre intensity). The only parameter that 
changes more considerably is the Lorentzian parameter $B$, but we note that this
procedure to fit the stray-light is known to be very insensitive to this parameter
(see Mart\'\i nez Pillet \cite{valentin}). The
stray-light levels far away from the solar limb are known to be more closely linked to the values
of the Lorentzian weighting parameter $M$. Indeed, it is a change in this parameter from $M=0.116$
in 1996
to $M=0.17$ in 2003 that explains the steady increase in scattered light in the MDI instrument
seen in Fig. 3.

\begin{table*}\caption{Spread function parameters from Mercury transit data}
\begin{tabular}{lcccc}\hline \hline
Parameter 	&Frame 1 UT 07:59:30	&Frame 2 UT 09:35:30	&Frame 3 UT 11:11:30	&Frame 4 UT 12:47:30 \\ \hline
$b_1$  (\arcsec) &      2            &      2          &      2       &   2         \\
$b_2$  (\arcsec) &      6            &      6          &      6       &   6         \\
$b_3$ (\arcsec)  &      18           &      17         &      16      &  17          \\
$B$  (\arcsec)   &      61           &      47         &      40      &  46          \\
$m_1$             &  0.676          & 0.672        &   0.693      & 0.694           \\
$m_2$             &  0.117          & 0.123        &   0.108      & 0.107           \\
$m_3$             &  0.035          & 0.033        &   0.029      & 0.029           \\
$M$               &  0.173          & 0.171        &   0.170      & 0.171           \\
\hline
\hline
\end{tabular}
\end{table*}
The first Gaussian, with a 2\arcsec\/ parameter (FWHM 3.5\arcsec), represents the pixel dominated 
MDI resolution. The other Gaussian and Lorentzian terms represent contributions from diffraction 
and scattering
by the optical components. Umbral scattered light is dominated by the widest Gaussian 
($b_3$, of around 30\arcsec\/ FWHM) 
and the Lorentzian  components. 

Once the spread function is derived for a given image, we proceed with its restoration. If 
$O(u,v)$ is the Fourier transform of the observed image, the restored image is computed as the inverse 
Fourier transform of $O^R(u,v)$, defined as follows:
\begin{equation}
O^R(u,v)=O(u,v)\Phi(q)T(q) 
\end{equation}
with $q=\sqrt{u^2+v^2}$ the radial coordinate in the Fourier domain. $\Phi(q)$ corresponds to the
division by the Fourier transform of the spread function, $t(q)$ (the modulation transfer function), 
defined within 
the optimum filter scheme. That is:
\begin{equation}
\Phi(q)={{t(q)}\over{t^2(q)+{{1}\over{SNR^2(q)}}}}
\end{equation}
with $SNR(q)$ the Signal-to-Noise Ratio of the data in the Fourier domain. The level of white noise 
used for this quantity was estimated, as usual, by observing the high frequency part of the power
spectrum of the data. Finally, $T(q)$ is a cosine based apodization window whose effect is to
reduce the restoration of the intermediate Gaussian component and completely avoid any restoration 
from the Gaussian component corresponding to the pixel sampling. The apodization window
leaves the restoration of the Lorentzian and broadest Gaussian component almost untouched. 
The exact level of apodization (48\%) was fine tuned by using the Mercury
observations. If a full restoration is used with no apodization window, umbral intensities
remain within 1-2 percent of their value obtained with apodization, 
but Gibbs oscillations are enhanced at the limb in the restored data. These Gibbs
oscillations also affect our penumbral intensities for the smallest spots and 
our apodization function helps to minimise their effect.

The restoration in the Fourier domain is made with images that are twice as
large as the original MDI images (i.e. 2048$\times$2048 frames). Only when these
extended images were used, could we recover a clean zero intensity value 
(in the restored frames) beyond
the solar limb (see Fig. 4). Restoration using normal size frames always
produced negative values in this area. The reason for this behaviour was found
to be the doubled sampling in the Fourier domain of the transform of
the Lorentzian component. This component, being the most extended one, has a
very sharp peak at low frequencies in the MTF. This peak was correctly sampled
only when the extended images were used. The stray light was artificially prolongated 
in the extended observed images using the fitted spread function parameters.

The observed core intensity of Mercury (12\arcsec\/ diameter) is 0.16 of the
surrounding photosphere and the mean value over the planet's disk 0.23. After the
restoration the core intensity is $-0.02$ (slight over correction) and the mean
value 0.09. 
This shows that our restoration provides core intensities that are correct to 
within a few percent and mean values that are under corrected by 10\% for pores 
that are of a size similar to Mercury. Sunspots of this size or larger
will have much better accuracies in both core and mean intensities.

A visual inspection of Fig. 10 shows that the correction applied by our algorithm 
turns out to be larger for larger sunspots and becomes smaller for the
smallest ones. At first sight this result is somewhat puzzling. One
expects smaller spots to show larger contamination levels (defined as observed minus
corrected intensities), but we caution that this
is only true as long as the intrinsic brightness of the spots remains the same for 
different sizes. A correlation between sunspot size and continuum
intensity (as found in this paper, i.e. increase in intensity with decreasing size) 
can mask this trend. To confirm that for small bright sunspots one can expect lower 
contamination levels, we performed simulations where a 
brightness-size relationship was assumed and estimated the contamination levels 
using the spread functions derived for the MDI instrument during the Mercury transit. 
The simulations consist of locating a sunspot of a known size and brightness at the 
centre of an artificial solar image that displays the CLV derived from our analysis. 
These images are convolved (also with 2048$\times$2048 frames) with the inferred spread 
function and the contamination level for this sunspot size is directly evaluated from 
the convolved image. The results can be seen in Fig. A.1. Three cases are presented, 
umbral intensity constant for all sizes (solid line), a linear decrease of umbral 
intensities with umbral radii (dashed line) and, finally, a steeper linear decrease 
for smaller umbrae with radii less than 8.6\arcsec and a constant intensity value for larger 
umbrae (dot-dashed line). The contamination levels obtained from the first case 
(constant intensity) show the expected result of monotonically decreasing values 
for larger umbrae (solid line). When the umbral intensity is smaller for larger umbrae 
(dashed line), the contamination level initially shows the same tendency of smaller levels for larger 
sunspots but soon changes the slope
and provides a larger contamination for larger sunspots. The 
third case shows that when there is a very strong dependence
between umbral intensities and sunspot sizes (smaller spots are brighter) the contamination
levels are always larger for larger sunspots. This tendency is reversed when the umbral
intensities remain constant for the largest sunspots. Then, the expected decrease of the
contamination levels for larger umbrae is found again. The shape of this (dot-dashed) 
curve is  reminiscent of the dependence between the observed contamination levels
and umbral sizes shown in Fig. 10. We thus
conclude that the observed increased brightness for smaller sunspots in this Figure 
is perfectly compatible with stray light corrected data and reflects a real tendency of 
solar spots.

\end{document}